\documentclass{fac}

\ifprodtf \else \usepackage{latexsym}\fi
\newif\ifwithappendix

\usepackage[svgnames,usenames]{xcolor}

\usepackage{multirow}

\usepackage{paralist}

\usepackage{setspace}
\usepackage{calc}

\usepackage{groove2tikz}
\usepackage{booktabs}

\usepackage{setspace}
\usepackage{graphicx}
\usepackage{rotating}
\usepackage{adjustbox}

\usepackage{listings}

\usepackage{url}

\usepackage{xspace}

\usepackage{pdfcomment}

\usepackage{amssymb}
\usepackage[normalem]{ulem}

\usepackage{listings,textcomp}

\lstdefinestyle{basestyle}
{%
	frame=none,
	tabsize=2,
	numbers=none,
	basicstyle=\footnotesize\ttfamily,
	upquote=true,
	columns=fixed,
	showstringspaces=false,
	numberstyle=\ttfamily\tiny,
	numbersep=5pt,
	extendedchars=true,
	breaklines=true,
	showtabs=false,
	showspaces=false,
	showstringspaces=false,
	identifierstyle=\ttfamily,
	keywordstyle=\ttfamily\bfseries\color[rgb]{0,0,0.6},
	keywordstyle=[2]\ttfamily\bfseries\color{DarkRed},
	commentstyle=\it\ttfamily\color[rgb]{0,0.4,0},
	stringstyle=\it\ttfamily\color[rgb]{0.4,0,0},
	captionpos=b,
  numberbychapter=false,
	escapeinside={(*@}{@*)}
}
\lstdefinestyle{eiffellistingstyle}
{%
	style = basestyle,
	basicstyle=\small\ttfamily,
	language = Eiffel,
	morekeywords = {across, alias, all, and, as, check, class, interface, creation, create, debug, deferred, do, else, elseif, end, ensure, expanded, export, external, False, feature, from, frozen, if, implies, indexing, infix, inherit, inspect, invariant, is, like, local, loop, not, obsolete, old, once, or, prefix, redefine, rename, require, rescue, retry, select, some, strip, then, True, undefine, unique, until, variant, when, xor, Result, Current, Void, attached, detachable, agent, separate},
	morecomment=[l]--
}

\lstdefinestyle{eiffelinlinestyle}
{%
	style=eiffellistingstyle,
	basicstyle=\normalsize\ttfamily
}

\lstnewenvironment{eiffelcode}[1][]
{%
	\lstset{style=eiffellistingstyle,#1}
}{}

\lstdefinestyle{controlprgstyle}
{%
	style = basestyle,
	basicstyle=\scriptsize\ttfamily,
	language = C,
	morekeywords = {alap, while, try},
	morekeywords=[2]{recipe},
}

\newcommand{\incode}[1]{\lstinline[style=eiffelinlinestyle]{#1}}

\usepackage{tikz}
\usetikzlibrary{calc,matrix,positioning,fit,arrows,automata,shapes.misc,shapes.geometric,shapes.arrows,shapes.callouts,shapes.symbols,backgrounds,decorations,decorations.pathmorphing}

\newcommand{\gts}{\textsc{Gts}\xspace}
\newcommand{\scoop}{\textsc{Scoop}\xspace}
\newcommand{\dscoop}{\textsc{\mbox{D-Scoop}}\xspace}
\newcommand{\groove}{\textsc{Groove}\xspace}

\newcommand{\msg}[1]{\smash{\tikz[baseline=(msgnode.base)]\node[inner sep=2pt,rounded corners=.5ex,draw,rectangle](msgnode){\scriptsize{#1}};}\xspace}

\newcommand{\scoopgts}{\textsc{Scoop}-\textsc{Gts}\xspace} 
\newcommand{\scoopgraph}{\scoop-Graph\xspace}
\newcommand{\scoopgraphs}{\scoop-Graphs\xspace}
\newcommand{\qoq}{\textsc{QoQ}\xspace}
\newcommand{\req}{\textsc{RQ}\xspace}

\newcommand{\fifo}{\textsc{Fifo}\xspace}

\newcommand{\model}[1]{%
\ifx&#1& \scoopgts
\else \mbox{\scoopgts}(#1)\xspace
\fi}

\newcommand{\cfg}{\textsc{Cfg}\xspace}

\newcommand{\KK}{\ensuremath{\mathbb{K}}\xspace}

\newcommand{\maude}{\textsl{Maude}\xspace}

\newcommand{\eg}{e.g.\ }
\newcommand{\ie}{i.e.\ }

\newcommand{\myparagraph}[1]{\paragraph{#1}}

\newcommand{\ruleE}{rule engine\xspace}
\newcommand{\confG}{configuration graph\xspace}
\newcommand{\confGs}{configuration graphs\xspace}

\newcommand{\ConfGs}{Configuration graphs\xspace}

\newcommand{\addcompolabel}[1]{%
  \draw (#1.north east) node[anchor=north east,minimum height=2ex,minimum width=1.5ex,draw,outer sep=3pt] (compolabel){};
  \draw[fill=white] (compolabel.west) ++(0.2,0.05) rectangle +(-0.2,0.05);
  \draw[fill=white] (compolabel.west) ++(0.2,-0.05) rectangle +(-0.2,-0.05);
 }

\newcommand{\pluginto}[3][0.5]{%
\draw  ($#2!#1!#3$) coordinate (plugintoanchor);
     \draw  (plugintoanchor) edge[)-] #2;
     \draw (plugintoanchor) edge[o-] #3;
}

\title[A semantics comparison workbench for a concurrent, asynchronous, distributed programming language]
      {A semantics comparison workbench for a concurrent, asynchronous, distributed programming language}

\author[C.~Corrodi, A.~Heu{\ss}ner, and C.M.~Poskitt]
    {Claudio Corrodi$^1$, Alexander Heu{\ss}ner$^2$, and Christopher M. Poskitt$^3$\\
     $^1$Software Composition Group, University of Bern, Switzerland\\
     $^2$Software Technologies Research Group, University of Bamberg, Germany\\
   $^3$Singapore University of Technology and Design, Singapore}

\correspond{C.~Corrodi (\url{corrodi@inf.unibe.ch}), A.~Heu{\ss}ner (\url{alexander.heussner@uni-bamberg.de}), or C.M.~Poskitt (\url{chris_poskitt@sutd.edu.sg}).}

\pubyear{2016}
\pagerange{\pageref{firstpage}--\pageref{lastpage}}

\begin{document}
\label{firstpage}

\makecorrespond

\maketitle

\begin{abstract}
  A number of high-level languages and libraries have been proposed that offer novel and simple to use abstractions for concurrent, asynchronous, and distributed programming. The execution models that realise them, however, often change over time---whether to improve performance, or to extend them to new language features---potentially affecting behavioural and safety properties of existing programs. This is exemplified by \scoop, a message-passing approach to concurrent object-oriented programming that has seen multiple changes proposed and implemented, with demonstrable consequences for an idiomatic usage of its core abstraction. We propose a  \emph{semantics comparison workbench} for \scoop with fully and semi-automatic tools for analysing and comparing the state spaces of programs with respect to different execution models or semantics. We demonstrate its use in checking the consistency of properties across semantics by applying it to a set of representative programs, and highlighting a deadlock-related discrepancy between the principal execution models of \scoop. Furthermore, we demonstrate the extensibility of the workbench by generalising the formalisation of an execution model to support recently proposed extensions for distributed programming. Our workbench is based on a modular and parameterisable graph transformation semantics implemented in the \groove tool. We discuss how graph transformations are leveraged to atomically model intricate language abstractions, how the visual yet algebraic nature of the model can be used to ascertain soundness, and highlight how the approach could be applied to similar languages.
\end{abstract}

\begin{keywords}
  concurrent asynchronous programming, distributed programming with message passing, operational semantics, runtime semantics, graph transformation systems, verification/analysis parameterised by semantics, concurrency abstractions, object-oriented programming, software engineering, \scoop, \groove
\end{keywords}

\section{Introduction}

In order to harness the power of modern architectures, software engineers must program with concurrency, asynchronicity, parallelism, and distribution in mind. This task, however, is fraught with difficulties: data races and deadlocks can result from the most subtle of errors in synchronisation code, and unexpected program behaviours can emerge from the interactions between processes. To address this, a number of novel programming APIs, libraries, and languages have been proposed that provide programmers with more intuitive models of concurrency and distribution, such as block-dispatching in Grand Central Dispatch~\cite{GCD-Reference}, ``sites'' and concurrency primitives in Orc~\cite{Kitchin-et_al09a}, or message-passing and active objects in languages such as \scoop~\cite{West-NM15b} and Creol~\cite{JohnsenOY06}.

The high-level programming abstractions that such languages provide rely on intricate implementations that must maximise concurrency and performance whilst ensuring that programs still behave as the programmer expects them to. Devising execution models that successfully reconcile these requirements, however, is challenging: a model too restrictive can deny desirable concurrency and cause unnecessary bottlenecks, but a model too permissive might lead to surprising and unintended program executions emerging. Furthermore, execution models evolve and change over time as language designers seek to improve performance, and seek to support new language constructs or applications. Comparing the performance of different execution models is as simple as benchmarking their implementations. It is much harder, however, to detect and analyse the subtle effects of semantic changes on behavioural or safety properties, which have the potential to affect existing programs written and tested under older execution models.

One language that clearly exemplifies these issues is \scoop~\cite{West-NM15b}, a message-passing approach to concurrent object-oriented programming. \scoop provides concurrency in a very shielded way, designed to allow programmers to introduce it while still maintaining the modes of reasoning they are familiar with from sequential programs, e.g.~localised pre- and postcondition reasoning, and interference-free method execution over multiple objects. The fundamental language abstractions of \scoop were informally proposed as early as the `90s~\cite{Meyer93a,Meyer97a}, but it took many more years to realise them effectively: multiple execution models~\cite{Brooke-PJ07a,Morandi-SNM13a,West-NM15b}, prototypes~\cite{Nienaltowski07a,Torshizi-OPC09a}, and production-level implementations~\cite{SCOOP-EiffelStudio-Reference} appeared over the last decade. Furthermore, the latest semantics is unlikely to be the last, as new language features continue to be proposed and integrated with the existing abstractions, e.g.~shared memory~\cite{Morandi-NM14a} and distributed programming extensions~\cite{Schill-PM16a}. Together, these can be seen as a family of semantics for the \scoop language, but a family that is partially-conflicting. To illustrate one such conflict, suppose that the following two blocks of code are being executed by two distinct and concurrent threads of control:

\begin{minipage}[t]{.5\textwidth}
	\vfill
 \begin{eiffelcode}
separate stack
do
	stack.push (1)
	stack.push (2)
	...
	stack.push (7)
end
\end{eiffelcode}
\end{minipage}
\hfill
\begin{minipage}[t]{0.5\textwidth}
\begin{eiffelcode}
separate stack
do
	stack.push (8)
	x := stack.top
end
\end{eiffelcode}
\end{minipage}
\hfill

\noindent Intuitively, the \incode{stack} object is some concurrent stack of integers, and each \incode{separate} block indicates to \scoop that the instructions within should be executed on the \incode{stack} in program text order, and without interference from the instructions of other threads. As a consequence, if the \incode{stack} is observed, it would be impossible for \incode{8} to appear anywhere in-between \incode{1} through to \incode{7}; furthermore, the value stored to \incode{x} will always be \incode{8}. The synchronisation to achieve this---which allows the programmer to reason about a \incode{separate} block as if it were sequential code---is the responsibility of \scoop, and generalises to blocks over multiple concurrent objects. The execution models orchestrating this, however, have changed over time: the original model~\cite{Morandi-SNM13a} had the effect of blocking concurrent objects for the full duration of \incode{separate} blocks (e.g.~blocking the \incode{stack} until \incode{stack.push(1)} through to \incode{stack.push(7)} are all requested), whereas the current model~\cite{West-NM15b} only blocks when necessary for the sake of performance (e.g.~competing \incode{stack.push} commands are logged simultaneously, but in a special nested queue structure that ensures the order guarantees).

While both execution models maintain the order guarantees, they can lead to the same program behaving quite differently: under the older model, for example, programs are more prone to deadlocking, whereas under the current model, existing programs that relied (whether intentionally or not) on the coarse-grained blocking as a lock on some resource may no longer work as the programmer intended. Despite such substantial changes, the different semantics of \scoop have only ever been studied in isolation: little has been done to formally \emph{compare} the execution traces permitted under different execution models and ensure that their behavioural and safety properties are consistent (in fact, they are not). While some comprehensive, tool-supported semantic formalisations do exist---in \maude's conditional rewriting logic~\cite{Morandi-SNM13a} for example, and in a custom-built \textsc{Csp} model checker~\cite{Brooke-PJ07a}---they are tied to particular execution models, do not operate on actual source code, and are geared towards ``testing'' the semantics as they are unable to scale to more general verification tasks. Owing to the need to handle waiting queues, locks, asynchronous remote calls, and several other intricate features of the \scoop execution models, these formalisations quickly become very complex, not only blowing up the state spaces that need to be explored, but also making it difficult to confirm their soundness with language designers---one of the few means of ascertaining soundness in the absence of precise documentation.

\myparagraph{Our Contributions.} We propose a \emph{semantics comparison workbench} for \scoop, with fully and semi-automatic tools for analysing and comparing the state spaces of programs with respect to different execution models or semantics. Our workbench is based on a graph transformation system (\gts) formalisation that:
\begin{inparaenum}[(i)]
  \item covers the principal concurrent asynchronous features of the language, using \gts rules and control programs (strategies) to atomically model their intricate effects on \scoop states (i.e.~on control flow, concurrent object structures, waiting queues, and locks);
  \item is modular, parameterisable, and extensible, allowing for \scoop's different semantics to re-use common components, and to seamlessly plug-in distinct ones (e.g.~for storage, control, synchronisation); and
  \item is implemented in the general-purpose \gts tool \groove~\cite{Ghamarian-MRZZ12a}, providing out-of-the-box state space analyses for comparing programs under different \scoop semantics.
\end{inparaenum}
We demonstrate the use of the workbench for checking the consistency of properties across semantics by applying it to a set of representative \scoop programs, and highlighting a deadlock-related discrepancy between the principal execution models of the language. Furthermore, we demonstrate the extensibility of the workbench by generalising one of the semantics to support the features of \dscoop~\cite{Schill-PM16a}, a prototype extension of \scoop for distributed programming. We discuss how the visual yet algebraic nature of our \gts models can be used to ascertain soundness, and highlight how our approach could be applied to similar concurrent, asynchronous, distributed languages.

This is a revised and extended version of our FASE~2016 paper, \emph{``A Graph-Based Semantics Workbench for Concurrent Asynchronous Programs''}~\cite{Corrodi-HP16a} (itself based upon the preliminary modelling ideas in~\cite{GaM2015}), adding the following new content:
\begin{inparaenum}[(i)]
	\item a new \gts semantics covering the distributed programming abstractions of \dscoop, formalised orthogonally to the others by extending an existing semantics and not just replacing the components of one;
  \item a presentation of the underlying, compositional metamodel to which the family of \scoop and \dscoop semantics all conform, including a discussion of the metamodel's genericity;
	\item an expanded evaluation that additionally explores the state spaces of our benchmarks in fully distributed contexts; and
  \item a significantly revised presentation, including new details, explanations, and examples throughout the paper.
\end{inparaenum}

For language designers, this paper presents a transferable approach for checking the consistency of concurrent asynchronous programs under competing language semantics. For the graph transformation community, it presents our experiences of applying a state-of-the-art \gts tool to a non-trivial and practical problem in programming language design. For the broader verification community, it highlights the need for verification parameterised by different semantics, and demonstrates how \gts-based formalisms and tools can be used to derive an effective, modular, and extensible solution. Finally, for software engineers, it provides a workbench for crystallising their mental models of \scoop, potentially helping them to write better quality code and understand how to port it across different \scoop implementations.

\myparagraph{Plan of the Paper.}
We begin with an overview the \scoop concurrency model, its two most established execution models, and its distributed extension (Section~\ref{sec:fac:scoop}), before presenting some necessary \gts preliminaries (Section~\ref{sec:primer_gts}). Following this, we introduce a graph-based semantics metamodel for \scoop, and show how to formalise different, parameterisable semantics that conform to it (Section~\ref{sec:fac:genericmodel}). We propose a formal \gts-based model in \groove for the family of \scoop semantics, and implement it in a small toolchain (Section~\ref{sec:fac:tool}), allowing us to compare the state spaces of representative \scoop programs under different semantics (Section~\ref{sec:fac:analysis}) and highlight a deadlock-related discrepancy. Finally, we examine some related work (Section~\ref{sec:fac:related_work}), and conclude with a summary of our contributions and some future research directions (Section~\ref{sec:fac:conclusion}).

\section{SCOOP: A Concurrent, Asynchronous, Distributed Programming Language}
\label{sec:fac:scoop}

\scoop~\cite{West-NM15b} is a message-passing approach to concurrent object-oriented programming that aims to preserve the well-understood modes of reasoning enjoyed by sequential programs, such as sequential consistency, interference-free method execution over multiple objects, and pre- and postcondition reasoning over blocks of code. In order to achieve this, it provides its users with concurrency abstractions that are easier to reason about than threads, minimal new language syntax, and an implementation responsible for orchestrating the synchronisation. While \scoop has been studied principally in the context of concurrency, its programming abstractions also generalise to distributed systems by means of an additional layer for coordinating requests over a network~\cite{Schill-PM16a}.

This section presents an overview of \scoop's most important features. First, we describe handlers, \emph{separate} objects, and \emph{separate} blocks---\scoop's main concurrency abstractions. We demonstrate the reasoning they allow programmers to do in some simple examples. Second, we compare the two most established execution models for orchestrating the synchronisation, and highlight the rationale for their different approaches. Finally, we discuss \dscoop~\cite{Schill-PM16a}, a prototype extension of \scoop that extends one particular execution model with support for distributed programming.

Throughout this paper we present \scoop with respect to the syntax and terminology of its principal implementation for Eiffel~\cite{SCOOP-EiffelStudio-Reference}. We remark that the ideas, however, can be implemented for any other object-oriented language (as explored, e.g.~for Java~\cite{Torshizi-OPC09a}).

\subsection{Language Abstractions and Execution Guarantees}

\myparagraph{Handlers and Separate Objects.} In \scoop, every object is associated with a \emph{handler} (also called a \emph{processor}), a concurrent thread of control with the exclusive right to call methods on the objects it handles. Object references may point to objects sharing the same handler (\emph{non-separate} objects) or to objects with distinct handlers (\emph{separate} objects). Method calls on non-separate objects are executed immediately by their shared handler. To make a call on a separate object, however, a \emph{request} must be sent to the distinct handler of that object. If the method requested is a \emph{command} (i.e.~it does not return a result), then it is executed asynchronously, leading to concurrency; if it is a \emph{query} (i.e.~a result is returned and must be waited for), then it is executed synchronously. Note that handlers cannot synchronise via shared memory: only by exchanging requests.

In \scoop, objects that may have different handlers are declared with a special \incode{separate} type. In order to request method calls on objects of \incode{separate} type, programmers simply make the calls within so-called \emph{separate blocks} (the type system prevents calls outside of such blocks). These can be declared explicitly (we will use the syntax \incode{separate x,y,}\ \dots\ \incode{do}\ \dots\ \incode{end}), but whenever a \incode{separate} object is a formal parameter of a method, the body of that method is implicitly a separate block too. The underlying implementation is then responsible for orchestrating the synchronisation between handlers implied by the separate blocks.

\myparagraph{Execution Guarantees.} \scoop provides strong guarantees about the execution of calls in separate blocks, in order to help programmers reason more ``sequentially'' about their concurrent code and avoid typical synchronisation bugs. In particular, within a separate block, requests for method calls on \incode{separate} objects are always logged by their handlers in the order that they are given in the program text; furthermore, there will never be any intervening requests logged from other handlers. These guarantees apply regardless of the number of \incode{separate} objects and handlers involved in a separate block. As a consequence, programmers can write code over multiple concurrent objects that: (1)~is guaranteed to be data race free; and (2)~can be reasoned about sequentially and independently of the rest of the program.

To illustrate, consider the following separate blocks (adapted from~\cite{West-NM15b}) that set the ``colours'' of two \incode{separate} objects, \incode{x} and \incode{y}. Suppose that a handler is about to enter the separate block to the left, and concurrently, a distinct handler is about to enter the separate block to the right:

\begin{minipage}[t]{.5\textwidth}
	\vfill
 \begin{eiffelcode}
separate x,y
do
	x.set_colour (Green)
	y.set_colour (Green)
end
\end{eiffelcode}
\end{minipage}
\hfill
\begin{minipage}[t]{0.5\textwidth}
\begin{eiffelcode}
separate x,y
do
	x.set_colour (Indigo)
	a_colour = x.get_colour
	y.set_colour (a_colour)
end
\end{eiffelcode}
\end{minipage}
\hfill

\noindent The two separate blocks contain a mix of commands and queries, which are issued as asynchronous and synchronous requests respectively to the handlers of \incode{x} and \incode{y}. The body of the leftmost separate block asynchronously sets the colours of \incode{x} and \incode{y} to be \incode{Green}. The body of the rightmost block first asynchronously sets \incode{x} to \incode{Indigo}, then synchronously queries the colour of \incode{x}, then asynchronously sets the colour of \incode{y} to the result of that query. The \scoop guarantees ensure that for the duration of a \incode{separate x,y} block, no intervening requests can be logged on the handlers of \incode{x} or \incode{y}. As a result, it would not be possible to observe the colours in an intermediate state: both of them would be observed as \incode{Green}, or both of them would be observed as \incode{Indigo}. Interleavings permitting any other combination of the colours are completely excluded. This additional control over the order in which concurrent requests are handled represents a twist on classic message-passing approaches, such as the actor model~\cite{Agha86a}, and programming languages like Erlang~\cite{Armstrong-VW96a} that implement them.

\myparagraph{Wait Conditions.} \scoop also provides a mechanism for synchronising on conditions, built on top of Eiffel's native support for contracts\footnote{Contracts can also be supported in other object-oriented languages, e.g.~via JML~\cite{Burdy_et-al05a} for Java, or Code Contracts~\cite{Code-Contracts} for C\#.}. In sequential Eiffel, preconditions (keyword \incode{require}) and postconditions (\incode{ensure}) express conditions on the state that should hold at the beginning and end of a method execution. They are executable queries and can be monitored at runtime. In \scoop, however, if a precondition involves \incode{separate} objects, then it is re-interpreted as a \emph{wait condition} that must be synchronised on. The body of the method (which is implicitly a separate block) is not entered until the condition becomes true. We remark that postconditions do not have a special concurrent re-interpretation, and are simply (optionally) logged as requests immediately upon exiting the method body.

Consider the following excerpt from a \scoop program solving the producer-consumer problem:

\begin{minipage}[t]{.47\textwidth}
	\vfill
 \begin{eiffelcode}
put_on_buffer (a_buffer: separate BOUNDED_BUFFER[INTEGER]; an_element: INTEGER)
	require
		not a_buffer.is_full
	do
		a_buffer.put (an_element)
	ensure
		not a_buffer.is_empty
		a_buffer.count = old a_buffer.count + 1
	end
\end{eiffelcode}
\end{minipage}
\hfill
\begin{minipage}[t]{0.47\textwidth}
\begin{eiffelcode}
remove_from_buffer (a_buffer: separate BOUNDED_BUFFER [INTEGER]): INTEGER
	require
		not a_buffer.is_empty
	do
		a_buffer.consume
		Result := a_buffer.last_consumed_item
	ensure
		a_buffer.count = old a_buffer.count - 1
	end
\end{eiffelcode}
\end{minipage}
\hfill

\noindent Here, a number of concurrently executing producers (left) and consumers (right) must respectively add and remove elements from a buffer of bounded size that has its own concurrent handler (the buffer is of \incode{separate} type). Producers must not attempt to add an element to the buffer when it is full; consumers must not attempt to remove an element from the buffer when it is empty. These requirements are expressed as wait conditions in the \incode{require} clauses of the respective methods; they must become true before entering the method bodies (which are implicitly separate blocks, since the buffer is a formal argument). In the case of producers, \scoop guarantees that the request to call \incode{a_buffer.put(an_element)} will be logged on the handler of \incode{a_buffer} in an order such that \incode{not a_buffer.is_full} is true when it is executed; similar for consumers with \incode{a_buffer.consume} and the wait condition \incode{not a_buffer.is_empty}. 

\subsection{Execution Models}

The programming abstractions of \scoop require an \emph{execution model} that specifies how requests between handlers should be processed. Two contrasting models have been supported by the \scoop implementation over its evolution: initially, an execution model we call Request Queues (\req)~\cite{Morandi-SNM13a}, and an execution model that has since replaced it which we call Queues of Queues (\qoq)~\cite{West-NM15b}. In the following we compare the two models and highlight a semantic discrepancy between them.

\myparagraph{Request Queues.} The \req execution model associates each handler with a single \fifo queue for storing incoming requests. To ensure the \scoop execution guarantees, each queue is protected by a lock. For a handler to log a request on the queue of another handler, the former must first acquire the lock protecting the latter. Once the lock is acquired, it can log requests on the queue without interruption.

Under the \req model, upon entering a \incode{separate x,y,...} block, the handler must simultaneously acquire locks on the request queues associated with the handlers of \incode{x,y,...} and must hold them for the duration of the block. This coarse-grained solution successfully prevents intervening requests from being logged, but leads to performance bottlenecks in several situations, e.g.~multiple handlers vying for the lock of a highly contested request queue.

\begin{figure}
	\centering
	\includegraphics[scale=0.4]{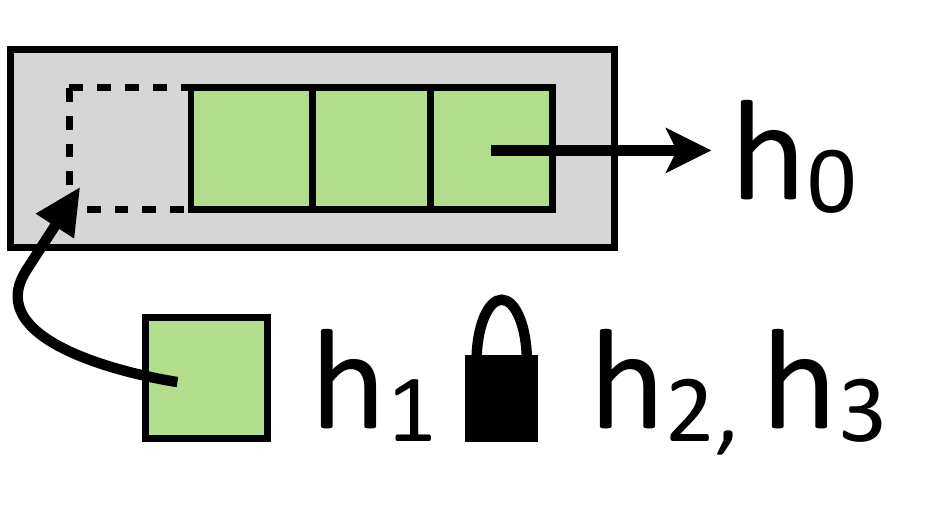}
  \vspace{-2ex}
	\caption{Three handlers ($h_1$, $h_2$, $h_3$) logging requests on another ($h_0$) under \req}\label{fig:em-rq-only}
\end{figure}

Figure~\ref{fig:em-rq-only} visualises three handlers ($h_1$, $h_2$, $h_3$) attempting to log requests (green squares) on the queue associated with another handler ($h_0$) under \req. Here, $h_1$ has obtained the lock (i.e.~entered a separate block involving objects handled by $h_0$) and is able to log its requests on the queue of $h_0$ without interruption. Once $h_1$ releases the lock (i.e.~exits the separate block), $h_2$ and $h_3$ will contend for the lock in order to log the requests that they need to.

\myparagraph{Queues of Queues.} In contrast, the \qoq execution model associates each handler with a ``queue of queues'', a \fifo queue itself containing (possibly several) \fifo subqueues for storing incoming requests. Each subqueue represents a ``private area'' for a particular handler to log its requests, in program text order, and without any interference from other handlers (since they have their own dedicated subqueues).

Under the \qoq model, upon entering a \incode{separate x,y,...} block, the handler is no longer required to fight for the exclusive right to log requests. Instead, dedicated subqueues are simultaneously prepared by the handlers of \incode{x,y,...} on which requests can be logged without interference for the duration of the block. Should another handler also need to log requests on \incode{x,y,...}, then another set of dedicated subqueues are prepared, and the requests can be logged on them concurrently. The \qoq model thus removes a potential performance bottleneck of \req, but is still able to ensure the \scoop reasoning guarantees by wholly processing the subqueues, one-by-one in the order that they were created, and processing the requests within each subqueue in the order that they were logged there.

\begin{figure}
	\centering
	\includegraphics[scale=0.4]{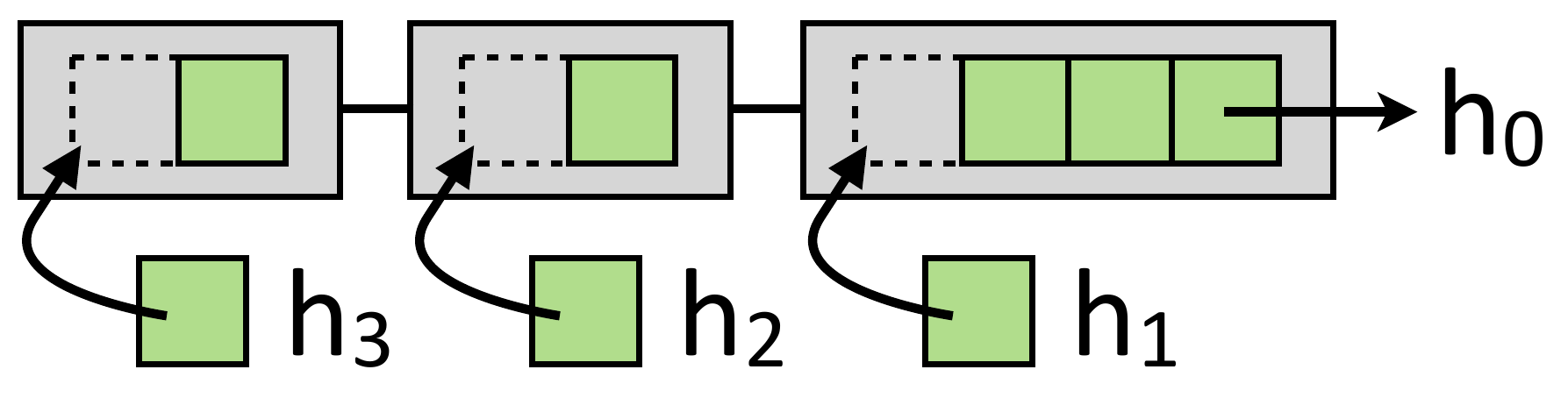}
  \vspace{-2ex}
	\caption{Three handlers ($h_1$, $h_2$, $h_3$) logging requests on another ($h_0$) under \qoq}\label{fig:em-qoq-only}
\end{figure}

Figure~\ref{fig:em-qoq-only} visualises three handlers ($h_1$, $h_2$, $h_3$) sending requests (green squares) to another handler ($h_0$) under \qoq. In contrast to \req (Figure~\ref{fig:em-rq-only}), the three handlers have access to dedicated subqueues and can log their requests concurrently. In highly asynchronous programs, this substantially reduces the amount of unnecessary blocking.

Note that the implementations of \req and \qoq (i.e.~compilers and runtimes) include a number of additional optimisations. While we do not model them in this paper, strictly speaking, these implementations could even be viewed as representing distinct semantics in the \scoop family.

\myparagraph{Semantic Discrepancies.} While both of the execution models correctly implement the programming abstractions of \scoop, discrepancies can arise in practice. We already highlighted that \qoq can lead to less blocking than \req, and thus boost performance. But it also has the potential to affect the intended functionality of a program. For example, in the mental model of programmers, separate blocks under \req had become synonymous with acquiring and holding locks. Such behaviour does not occur when entering separate blocks in \qoq.

This discrepancy is illustrated by the dining philosophers problem, in which concurrent processes (philosophers) must repeatedly acquire sole use of shared resources (forks) without causing a cyclic deadlock. The \scoop solution, provided as part of the official documentation~\cite{SCOOP-EiffelStudio-Reference}, attempts to solve it by representing philosophers and forks as separate objects---each with their own handlers---and modelling the acquisition of forks (eating) as a separate block. Consider Listing~\ref{listing:scoopdp2} and Listing~\ref{listing:scoopdp1}, which respectively provide the official solution and a variant with nesting. Under \req, Listing~\ref{listing:scoopdp2} (``eager'' philosophers) solves the problem by relying on the implicit simultaneous acquisition of locks on the forks' handlers; no two adjacent philosophers can be in their separate blocks at the same time. Under \req, Listing~\ref{listing:scoopdp1} (``lazy'' philosophers) can lead to a circular deadlock, since the philosophers acquire the locks in turn. Under \qoq however, neither version will deadlock, but neither version actually represents a solution: since all the requests are asynchronous, no blocking occurs at all, and philosophers can ``eat'' regardless of the states of other philosophers. While not a solution to dining philosophers under \qoq, the basic execution guarantees of \scoop remain satisfied. 

\begin{figure}
  \centering
  \begin{minipage}[t]{0.4\textwidth}
  \begin{eiffelcode}[caption={Eager philosophers},label=listing:scoopdp2]
separate left_fork, right_fork
do
  left_fork.use
  right_fork.use
end
(*@\vspace{5ex}@*)
  \end{eiffelcode}
  \end{minipage}
  \hspace{0.5cm}
  \begin{minipage}[t]{0.4\textwidth}
    \begin{eiffelcode}[caption={Lazy philosophers},label=listing:scoopdp1]
separate left_fork
do
  separate right_fork
  do
    left_fork.use
    right_fork.use
  end
end
    \end{eiffelcode}
  \end{minipage}
  \vspace{-2ex}
\end{figure}

\subsection{Extension for Distributed Programming}

Yet another competing semantics for \scoop is \dscoop (for Distributed \scoop)~\cite{Schill-PM16a}, which adds support for programming with separate objects over networks. Rather than replace an existing semantics, \dscoop extends the \qoq model: it aims to retain the abstractions, guarantees, and behaviours of \scoop under \qoq, while generalising them to distributed objects via an additional communication layer that remains hidden from the programmer. In the following, we provide an overview of how \scoop systems communicate over a network, and discuss a simple example.

\myparagraph{\dscoop.} In \dscoop, an instance of a running \scoop program (under \qoq) is called a \emph{node}\footnote{A \scoop program can be viewed as a \dscoop program with only one node.}. A node can open a connection to another node through a network socket, which is then shared by all of its handlers. Nodes communicate, via these connections, using \emph{messages} and \emph{replies}. Messages are sent from a \emph{client} node to a \emph{supplier}; replies are sent back from the supplier to the client to indicate the outcome.

When entering a separate block, it is now possible that one or more of the involved separate objects are handled on one or more remote nodes. For this new case, \dscoop introduces a (two-phase) locking protocol to allow for remote calls to be logged in a way that maintains the separate block guarantees, while minimising blocking as much as possible. The protocol involves three stages: (i)~a \emph{prelock stage}, for setting up remote subqueues in a correct order; (ii)~an \emph{issuing stage}, for logging object calls on those subqueues without interruption; and (iii)~an \emph{execution stage}, for dequeueing and executing those calls.

\myparagraph{Locking Protocol.} The prelock stage ensures the creation of subqueues across multiple remote handlers without interference. First, messages (with the subject \msg{PRELOCK}) are sent to the nodes of remote objects to announce that a handler in the client node wishes to enter a separate block that involves them. These are sent one-at-a-time and in a fixed order based on node IDs (to avoid deadlock). If a supplier receives a \msg{PRELOCK} message but is already involved in the prelock stage of another node, the client blocks. Once the supplier is available, it replies \msg{OK}, indicating that the client can unblock and send a \msg{PRELOCK} message to the next node involved (if any). Once these messages are all acknowledged, the client sends a \msg{LOCK} message to each ``prelocked'' supplier, which instructs them to prepare a private subqueue on the appropriate handler. By once again replying \msg{OK}, the suppliers are indicating that they are ready to receive and enqueue requests, and the prelock stage is over.

The issuing and execution stages are more straightforward, and typically overlap (in fact they must overlap if synchronous queries are involved). The client simply issues remote requests over the network as asynchronous \msg{CALL} or synchronous \msg{QCALL} messages, corresponding respectively to commands and queries. For the former, the supplier replies \msg{OK} as soon as the command is enqueued; for the latter, the supplier replies \msg{OK} as soon as the query is executed, and also returns the result.

When a supplier is involved in the prelock stage of a particular client, any other clients that try to involve it in a prelock stage are blocked. This blocking is crucial to ensure that subqueues are created without interference. Instead of blocking a competing node for the whole of a separate block, however, blocking only occurs during the prelock stage (i.e.~while subqueues are being set up); competing issuing stages can otherwise run concurrently. This allows for \dscoop systems to remain efficient, while lifting the execution guarantees and behaviours of \qoq to a distributed setting.

Additional details about the locking protocol as well as some example message-\&-reply exchanges are provided in~\cite{Schill-PM16a}. Note that \dscoop also provides some advanced mechanisms for recovering from failure (e.g.~compensation) which we do not explore in this paper.

\myparagraph{Distributed Example.} Consider the following code excerpt from a \dscoop implementation of a bank account management system:

\begin{eiffelcode}
transfer (source, target: separate ACCOUNT; amount: NATURAL)
do
	if source.balance >= amount then
		source.set_balance (source.balance - amount)
		target.set_balance (target.balance + amount)
	else
		-- Notify user (not shown)
	end
end
\end{eiffelcode}

\noindent The \incode{transfer} method allows some client to transfer an \incode{amount} of money from a \incode{source} account to a \incode{target} account. As the two bank accounts are of \incode{separate} type and provided as formal arguments, the body of \incode{transfer} implicitly forms a separate block.

Suppose that \incode{source} and \incode{target} are handled on two different remote nodes. Upon calling \mbox{\incode{transfer},} the client must follow the aforementioned locking protocol before it can enter the method body and start issuing requests. First, the client node sends a \msg{PRELOCK} message to the node containing \mbox{\incode{source}.} If (or when) the node is not being prelocked by another client, it replies \msg{OK}; the client then sends a \msg{PRELOCK} message to the node containing \incode{target} and waits for an \msg{OK}. At this stage, no other client can prelock the nodes containing the two accounts, i.e.~no other nodes can interrupt the process of generating subqueues. The client issues \msg{LOCK} requests to the suppliers, which trigger the creation of subqueues on the handlers of \incode{source} and \incode{target}. After both reply \msg{OK}, the client enters the body of \incode{transfer} (and the nodes of \incode{source} and \incode{target} become free to be prelocked by others). The \incode{balance} requests are issued as synchronous \msg{QCALL} messages, with the requests enqueued by the suppliers and waited for; the \incode{set_balance} commands are enqueued as asynchronous \msg{CALL} messages, and the client can exit the block before the final one (\mbox{\incode{target.set_balance}}) is executed. The requests are issued and executed without any interference.

Note that a (single node) concurrent version of \incode{transfer} would look exactly the same as this distributed version. The only difference emerges upon execution: if \incode{source} or \incode{target} are remote, then the implementation must follow the locking protocol. This is invisible to the programmer, who works with the same abstractions and execution guarantees regardless of where any \incode{separate} objects are actually located.


\section{Graph Transformation System Preliminaries}
\label{sec:primer_gts}

Our semantics comparison workbench for \scoop is based on graph transformation systems (\gts)---also known as graph-rewriting systems or graph grammars. Graph transformation is a flexible formalism for dynamic systems, and is well-suited to modelling the structures and relations in \scoop states (e.g.~object references, handlers, waiting queues). \gts is an inherently visual modelling approach, but is also anchored to a formal, algebraic basis, and is supported by a variety of practical tools.

In this section, we present a brief and informal overview the basic concepts of \gts, which should be sufficient to follow the rest of the paper. For a deeper introduction to \gts, we refer to reader to some standard textbooks, e.g.~\cite{Rozenberg:1997,Ehrig:2006}.

\myparagraph{Graph Transformation Systems.} \gts are rule-based systems for manipulating graphs. They can be seen as a computation abstraction, in which states (or configurations) are graphs\footnote{In this paper, our graphs are directed and labelled, with parallel edges allowed.}, and computational steps are rules that rewrite these graphs. In the state space of a \gts applied to some initial configuration, the states are thus graphs, and the transitions are rule applications.

\gts rules nondeterministically match a structural pattern in a graph and rewrite it. Rules consist of combinations of the following:
\begin{inparaenum}[(i)]
  \item a ``context'' in the graph that needs to be matched by the rule but is unchanged by it;
  \item a set of edges and nodes (and labels) that are added in this context;
  \item a context that is matched but removed by the rule;
  \item a negative application condition, \ie a part of the graph that when present, prohibits the application of the rule.
\end{inparaenum}

Typically, the rules of a \gts are applied to graphs nondeterministically (both in choosing the rule and choosing the match) and for as long possible. If a \gts consists of rules that unconditionally add new nodes or edges, for example, then it will be associated with an infinite transition system containing graphs of unbounded size. Many tools (e.g.~\groove~\cite{Ghamarian-MRZZ12a}, \textsc{GP 2}~\cite{Plump12a}) allow for control programs (or strategies) to be defined over the rules, adding a finer degree of control.

\myparagraph{Notation \& Example.} There are several ways to denote a \gts rule, but for simplicity, we will use the notation of \groove (since we use the tool in our workbench). In the following, we illustrate the application of \gts rules using a simple and intuitive example. Formally, rule applications are described in a proper categorical setting, via graph morphisms and pushout constructions (we refer the reader to~\cite{Ehrig:2006}).

Suppose we are using a graph to model a simple \textsc{Fifo} queue. Let us distinguish two types of nodes in our graph: a node labelled \smash{\tikz[baseline=(n.base)] \draw
node[draw,inner sep=1pt](n){Queue};} modelling the ``anchor'' of a queue, and message nodes \smash{\tikz[baseline=(n.base)]
\draw node[draw,inner sep=1pt](n){Message:a};} and
\smash{\tikz[baseline=(n.base)] \draw node[draw,inner
sep=1pt](n){Message:b};} labelled with the \tikz[baseline=(n.base)] \draw node (n){Message}; type and some contents, either \tikz[baseline=(n.base)] \draw node (n){a}; or \tikz[baseline=(n.base)] \draw node (n){b};. Furthermore, let us distinguish edges labelled with ``next'', representing pointers towards the tail of the \textsc{Fifo} queue.

Consider the following \gts rules \texttt{Append\_a} and \texttt{Append\_b}:\\
\begin{tikzpicture}
  \draw node[draw] (1) {Queue};
  \draw (4,0) node[draw] (2) {};
  \draw (2) edge[eraser_edge] node[fill=white] {next} (1);
  \draw (2,-1) node[creator_node] (3) {Message:a};
  \draw (3) edge[creator_edge] node[fill=white] {next} (1);
  \draw (2) edge[creator_edge] node[fill=white] {next}(3);

  \draw (-1,-0.5) node[anchor=east] {\tt Append\_a:};

  \begin{scope}[xshift=8cm]
    \draw node[draw] (1) {Queue};
    \draw (4,0) node[draw] (2) {};
    \draw (2) edge[eraser_edge] node[fill=white] {next} (1);
    \draw (2,-1) node[creator_node] (3) {Message:b};
    \draw (3) edge[creator_edge] node[fill=white] {next} (1);
    \draw (2) edge[creator_edge] node[fill=white] {next}(3);

    \draw (1.3,0.7) node[draw,embargo_node] (nac1) {Message};
    \draw (4,0.7) node[draw,embargo_node] (nac2) {Message};

    \draw[embargo_edge] (nac1) -| node[fill=white]{next}(1);
    \draw[embargo_edge] (nac2) -- node[fill=white]{next$^3$}(nac1);

    \draw (-1,-0.5) node[anchor=east] {\tt Append\_b:};
  \end{scope}

\end{tikzpicture}

\noindent Following \groove's notation, solid black nodes and edges are matched by the rule (but not deleted), dashed blue nodes and edges are matched and deleted, green ones are newly created, and red ones must not be present. Intuitively, the application of a rule to a graph is a three-step procedure: first, the black and blue structure is matched in a context where the red structure is not present; second, the blue structure in the match is deleted; finally, the green structure is created. Thus, \texttt{Append\_a} deletes a ``next'' edge incident to the anchor and inserts a \tikz[baseline=(n.base)] \draw node (n){Message:a}; node in its place; \texttt{Append\_b} does the same for a \tikz[baseline=(n.base)] \draw node (n){Message:b}; node, but only if the \textsc{Fifo} queue has at most three elements.

Suppose that we have the following initial configuration, which models an empty \textsc{Fifo} queue:

  \begin{tikzpicture}
    \draw node[draw] (n) {Queue};
    \draw (n) edge[->,out=-30,in=30,looseness=8] node[fill=white](en){next}(n);
	\end{tikzpicture}

\noindent The application of \texttt{Append\_a} to this graph proceeds as follows:

  \begin{tikzpicture}
    \draw node[draw] (n) {Queue};
    \draw (n) edge[->,out=-30,in=30,looseness=8] node[fill=white](en){next} (n);

  \begin{scope}[yshift = -1.5cm,xshift=0.5cm]
    \draw node[draw] (1) {Queue};
    \draw (4,0) node[draw] (2) {};
    \draw (2) edge[eraser_edge] node[fill=white] (e1) {next} (1);
    \draw (2,-1) node[creator_node] (3) {Message:a};
    \draw (3) edge[creator_edge] node[fill=white] (e2) {next} (1);
    \draw (2) edge[creator_edge] node[fill=white] (e3) {next}(3);

  \end{scope}

  \begin{scope}[ultra thick,gray,loosely dotted]
    \draw (n) -- (1);
    \draw (n) -- (2);
    \draw (en) -- (e1);
  \end{scope}

  \begin{scope}[xshift=6cm]
    \draw node[draw] (nn) {Queue};
  \end{scope}

  \begin{scope}[yshift = -1.5cm,xshift=8cm]
    \draw node[draw] (1) {Queue};
    \draw (4,0) node[draw] (2) {};
    \draw (2) edge[eraser_edge] node[fill=white] (e1) {next} (1);
    \draw (2,-1) node[creator_node] (3) {Message:a};
    \draw (3) edge[creator_edge] node[fill=white] (e2) {next} (1);
    \draw (2) edge[creator_edge] node[fill=white] (e3) {next}(3);

  \end{scope}

  \begin{scope}[xshift=12cm]

  \draw node[draw] (nn) {Queue};
  \draw (2,0) node[draw] (nma) {Message:a};
  \draw (nn) edge[<-,bend left=30] node[fill=white](nn1){next} (nma);
  \draw (nma) edge[<-,bend left=30]node[fill=white](nn2){next} (nn);

  \end{scope}

  \begin{scope}[ultra thick,gray,loosely dotted]
    \draw (e2) edge[out=5,in=-110] (nn1);
    \draw (e3) -| (nn2);
    \draw (3) -| (nma);
  \end{scope}

  \draw (4,0) node {\bfseries \large $\Rightarrow$};
  \draw (8,0) node {\bfseries \large $\Rightarrow$};

  \draw[ultra thick, orange, dashed] (-1,.5) -- ++(4,0) -- ++(2,-2) -- ++(0,-2) -| (-1,.5);

  \draw[ultra thick, orange, dashed] (15,-3.5) -- (15,0.5) -- ++(-4,0) -- ++(-2,-1.5) -- ++(-2,0) |- (15,-3.5);

  \draw (2.5,-3.5) node[anchor=south] {\bfseries 1. match};
  \draw (6,-3.5) node[anchor=south] {\bfseries 2. delete};
  \draw (11,-3.5) node[anchor=south] {\bfseries 3. create};

  \end{tikzpicture}

\noindent The match of the rule is indicated by dashed grey lines. Observe that in this case, two of the nodes in the rule are mapped to the same node (i.e.~the match is non-injective).

  The rule \texttt{Append\_b} could have been applied to the same initial graph, too, since its negative application (
  \begin{tikzpicture}[baseline=(q.base)]
    \begin{scope}[inner sep=1pt]
      \draw node[draw] (q) {Queue};
      \draw (2.8,0) node[draw,embargo_node] (m1) {Message};
      \draw (6,0) node[draw,embargo_node] (m2) {Message};
      \draw[embargo_edge] (m1) -- node[fill=white]{next}(q);
      \draw[embargo_edge] (m2) -- node[fill=white]{next$^3$} (m1);
  \end{scope}
\end{tikzpicture}
) has no match. This specifies that there must not exist a message node reachable from the anchor node by four ``next'' edges. Observe that the negative application condition is only matching the \tikz[baseline=(n.base)] \draw node (n) {Message}; type of the labels (and not its contents, \tikz[baseline=(n.base)] \draw node (n) {a}; or \tikz[baseline=(n.base)] \draw node (n) {b};), and that a simple regular expression is used over the edges. As a consequence, \texttt{Append\_b} can only be applied if the \textsc{Fifo} queue has at most three elements.

After multiple applications of \texttt{Append\_a} and \texttt{Append\_b}, we can eventually derive a graph representing a \textsc{Fifo} queue that contains the message $abba$:

  \begin{tikzpicture}
    \draw (-2,0) node (dummy) {};
    \draw node[draw] (1) {Queue};
    \draw (3,0) node[draw] (2) {Message:a};
    \draw (6,0) node[draw] (3) {Message:b};
    \draw (9,0) node[draw] (4) {Message:b};
    \draw (12,0) node[draw] (5) {Message:a};

    \draw (2) edge[->] node[fill=white]{next} (1);
    \draw (3) edge[->] node[fill=white]{next} (2);
    \draw (4) edge[->] node[fill=white]{next} (3);
    \draw (5) edge[->] node[fill=white]{next} (4);
    \draw[<-] (5) -- ++(0,-0.4) -| node[pos=0.3,fill=white] {next} (1);
  \end{tikzpicture}

  We could extend the \gts, for example, with similar rules that remove messages from the queue, taking different actions depending on the contents.

\myparagraph{Control Programs.}  In general, a \gts tries to apply its rules in a nondeterministic fashion.
  More fine-grained control over the application of rules is possible with the help of control programs (also know as strategies) that specify in a declarative way how rules are to be applied. For example, the control program
  \texttt{alap Append\_b; Append\_a;} would apply the rule \texttt{Append\_b} as long as possible (\texttt{alap}) and then \texttt{Append\_a} once. This will always lead to a message queue containing $bbbba$.

  In \groove, control programs can also specify so-called recipes, which wrap functions over (possibly multiple) rules into a single transition. Control programs and recipes provide an ideal base for defining \gts in a modular way, \eg by supporting different implementations of recipes for different semantics of components (e.g.~different queuing models in our case).

  We refer the interested reader to~\cite{Ghamarian-MRZZ12a} and the documentation of the \groove tool for more details on control programs, recipes, and additional features of \gts rules, \eg $!=$ and $==$ edges (for explicitly expressing whether two matched nodes are distinct or not), or nested rules for matching universally ($\forall$) and existentially ($\exists$) quantified substructures.

\section{A Graph-Based Semantics Metamodel}
\label{sec:fac:genericmodel}
With the example of \scoop, we have motivated the need for a semantics comparison workbench that:
\begin{inparaenum}[(i)]
  \item can model features such as asynchronous remote calls and waiting queues;
  \item is modular (e.g.~for replacing \req synchronisation with \qoq) and extensible (e.g.~for lifting \qoq to \dscoop); and
  \item provides formal and automatic analyses for checking the consistency of behavioural and safety properties of programs under different semantics.
\end{inparaenum}
The following three sections present how we achieve this through the use of a \gts semantics, formalised in the \groove tool, and supported by a wrapper and simple toolchain.

In this section, we describe the first step of our process, in which we derive a graph-based, compositional metamodel to which the family of \scoop semantics (and possibly other message-passing language semantics) all belong. We use the metamodel to formally structure the sub-components of semantics and the interfaces between them. This abstract structure provides the basis of our approach to semantics parameterisation, which allow for common semantic components to be re-used across different execution models, and other semantic components to be plugged-in.

\myparagraph{Metamodel Overview.} We present a graph-based \emph{semantics metamodel} covering \scoop and other actor-like programming languages based on:
\begin{inparaenum}[(i)]
  \item message-passing concurrency;
  \item (active) objects, which are explicitly assigned to a handler and can only be accessed via this handler;
  \item asynchronous and synchronous calls between objects across different handlers; and
  \item different distribution topologies (i.e.~different ways of connecting or relating distributed runtimes).
\end{inparaenum}
The metamodel describes the structure of the semantic components found in \req, \qoq, and \dscoop, but abstracts away from concrete details (e.g.~computations on non-separate, separate, or remote objects). Different models expressing these details, however, can be \emph{plugged in} to the metamodel because of its compositionality, so long as they conform to the abstract boundaries and interfaces it defines. Semantic plug-ins of interest include, for example, different storage models, different queuing semantics, and different distribution topologies.

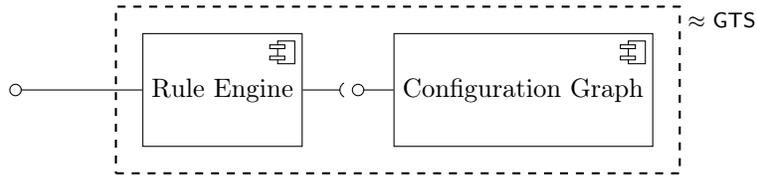
\begin{figure}\centering
   \begin{tikzpicture}
     \begin{scope}[every node/.style={draw,minimum height=1.5cm}]
       \draw (3,0) node (rte) {Rule Engine};
       \draw (7,0) node (rtg) {Configuration Graph};
       \draw[dashed,thick] node[fit=(rte)(rtg),inner sep=10pt] (l){};
     \end{scope}
     \path (0,0) node (x) {};
     \draw[->] (x) edge[o-] (rte);
     \path ($(rte.east)!.5!(rtg.west)$) coordinate (x);
     \draw (x)++(-0.05,0) edge[)-] (rte);
     \draw (x)++(0.05,0) edge[o-] (rtg);
     \draw (l.north east) node[anchor=north west]{$\approx$ GTS};
     \addcompolabel{rte}
     \addcompolabel{rtg}
   \end{tikzpicture}
   \caption{Overview of the \gts underpinning a semantics comparison workbench}

   \label{fig:mm_overview}
 \end{figure}

Our semantics metamodel describes the structure of a \gts, consisting of a \emph{\ruleE} and \emph{configuration graph} (Figure~\ref{fig:mm_overview}). The \ruleE encodes the step-wise operational rules of the semantics, whereas the \emph{configuration graph} encodes a snapshot of the state of the handlers, their objects, their current synchronisation topology, as well as a representation of the original \scoop program's control flow. Before launching a state-space exploration on a \scoop program, the \ruleE and corresponding \confGs must be initialised accordingly. The \ruleE can be parameterised by plugging in semantic components to simulate different ways of synchronising, queuing, and handling distributed objects (the choices of which are then reflected in the typing of the \confG). The \confG must also be initialised to encode the control-flow information of the original \scoop source code, as well as any expected initial configurations of handlers, objects, and topology. Our workbench is thus a front end responsible for initialising both aspects of the \gts (with respect to an execution model and \scoop program), and then launching simulations or analyses of the system's behaviour.

In the following, we present these building blocks of our metamodel in more detail, and demonstrate a semantic plug-in for storage. Without loss of generality, we assume---for simplicity of presentation---that objects are dynamically generated from flat class templates, \ie inheritance is flattened in a pre-compilation step. Thus, our (\scoop) objects consist of a finite fixed set of initialised variables with values that can be changed in program executions.

\myparagraph{Configuration Graphs.} \ConfGs, which encode snapshots of the states of handlers, are the heart of our metamodel, and describe the structure of \emph{configurations}\footnote{We will use \emph{configurations} and \emph{configuration graphs} synonymously.} in traces of rule applications. Each configuration encodes both static control flow information (extracted from the original program), as well as the dynamic states of the handlers, any objects under them, and the topologies that connect them.

\newcommand{\addstereotype}[2][Handler]{%
    \draw (#2.north west) node[anchor=north west,font=\tiny] {\guillemotleft #1\guillemotright};
}

\begin{figure}
  \centering
  \begin{tikzpicture}
    \begin{scope}[every node/.style={draw,minimum height=1.2cm}]
      \draw node (h1) {Handler 1};
      \draw (5,-.5) node (ta) {Topology Abstraction};
      \draw (10,0) node (hn) {Handler n};
      \draw (2,1) node (h2) {Handler 2};
      \draw (8,1) node (hnmo) {Handler n-1};
      \draw (5,-1.8) node[minimum width=13cm,minimum height=0.7cm] (cfg) {Control Flow Information};
    \end{scope}
    \draw (5,1) node {\textbf{\huge\dots}};

    \foreach \x in {h1,h2,hnmo,hn}{%
      \addcompolabel{\x}
      \addstereotype{\x}
    }

    \addcompolabel{cfg}
    \addcompolabel{ta}

    \pluginto{(h1)}{(ta)}
    \pluginto[0.45]{(h2)}{(ta)}
    \pluginto[0.45]{(hnmo)}{(ta)}
    \pluginto{(hn)}{(ta)}

    \pluginto{(cfg.north-|h1.south)}{(h1.south)}
    \pluginto[0.25]{(cfg.north-|h2.south)}{(h2.south)}
    \pluginto[0.25]{(cfg.north-|hnmo.south)}{(hnmo.south)}
    \pluginto{(cfg.north-|hn.south)}{(hn.south)}

    \draw[] node[draw,fit=(h1)(h2)(hn)(cfg),inner sep=10pt] (l){};
    \addcompolabel{l}

    \draw (l.north) node[anchor=north,font=] {\confG};

  \end{tikzpicture}
  \caption{Structure of a \confG (with $n$ handlers)}\label{fig:mm_configgraphs}
\end{figure}
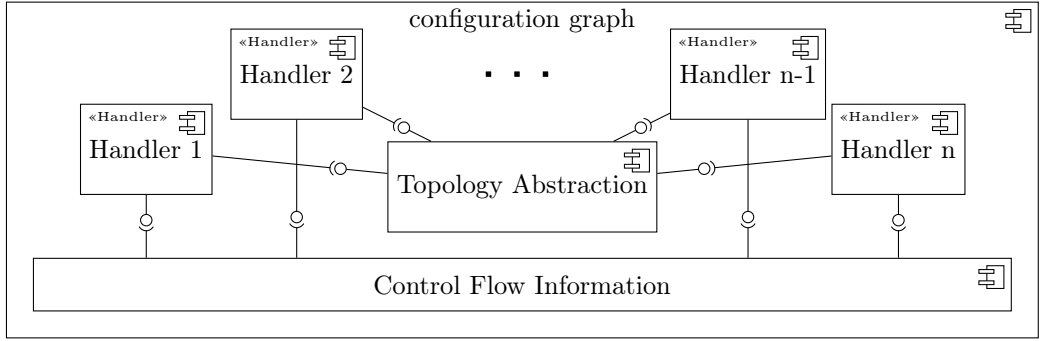

Figure~\ref{fig:mm_configgraphs} depicts the three principal components of \confGs and their connections. Each \emph{handler} defines an autonomous execution unit with exclusive access to some region of storage, and the ability to administer inter-handler synchronisation via the topology abstraction. Note that there is an a priori unbounded number of handler instances in a \confG. The \emph{topology abstraction} connects the handlers and defines the channels by which they synchronise. It also encodes name resolution and references between objects residing under the control of different handlers. Furthermore, it can be used to encode nodes and distributed communication channels. The \emph{control flow information} encodes the control flow graph of the program, including all information necessary to dynamically generate new objects, e.g.~via class templates. Note that the control flow information in a \confG is static and does not change under the execution of the rule engine.

\begin{figure}
  \centering
  \begin{tikzpicture}
    \begin{scope}[every node/.style={draw,minimum height=1.2cm}]
      \draw node[minimum height=0.8cm] (ctrl) {Control};
      \draw (0,1.7) node (storage) {Storage};
      \draw (3.7,0) node (sync) {Synchronisation};
    \end{scope}

    \addcompolabel{storage}
    \addcompolabel{sync}

    \path (9,0) node (ta) {topology abstraction};
    \path (0,-2) node (cfg) {control flow graph};

    \pluginto{(ctrl.north)}{(storage.south)}

    \pluginto{(ctrl.east)}{(sync.west)}

    \draw (ctrl) edge[-o] (cfg);
    \draw (sync) edge[-(] (ta);

    \draw[] node[draw,fit=(ctrl)(storage)(sync),inner sep=10pt] (l){};
    \draw (l.north) node[anchor=north](x) {\guillemotleft Handler\guillemotright};
    \addcompolabel{l}

    \draw (l.east|-sync.west) node[fill=white,rectangle,draw] (portsync) {};
    \draw (l.south-|ctrl.south) node[fill=white,rectangle,draw] (portcfg) {};


  \end{tikzpicture}
  \caption{Structure of a handler and its connections to its environment}

  \label{fig:mm_handler}
\end{figure}
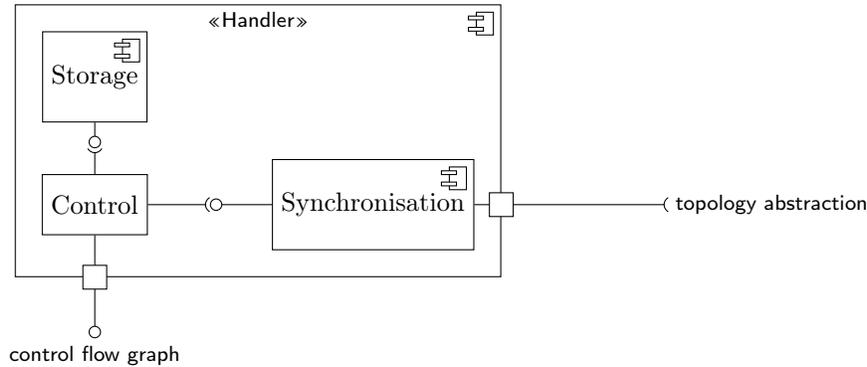

Our metamodel further divides each handler into three semantic subcomponents:
\begin{inparaenum}[(i)]
  \item  its \emph{control state}, recording the handler's current position in the control flow information;
  \item  its \emph{storage stage}, including a stack for recursion, and a heap containing objects, possibly with references to separate objects (i.e.~under the control of other handlers); and
  \item  a \emph{synchronisation} component connecting to the topology abstraction and including, for example, a dispatcher for outgoing requests and an input queue for storing requests received from other handlers.
\end{inparaenum}
These three subcomponents are depicted in Figure~\ref{fig:mm_handler}.

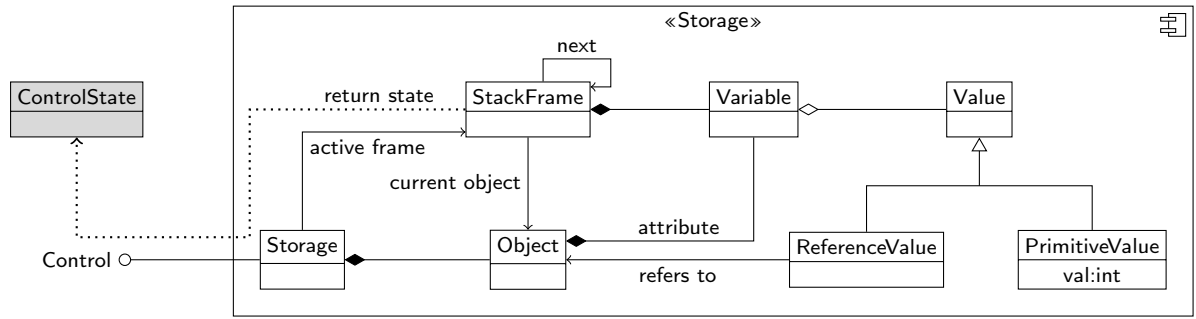
\begin{figure}
  \begin{tikzpicture}
    [class/.style={ draw, rectangle split, rectangle split parts=2}]

    \draw node[class] (s) {Storage};
    \draw (3,0) node[class] (obj) {Object};

    \draw (3,2) node[class] (sf) {StackFrame};

    \draw (6,2) node[class] (var) {Variable};

    \draw (9,2) node[class] (val) {Value};

    \draw (7.5,0) node[class] (refval) {ReferenceValue};

    \draw (10.5,0) node[class] (primval) {PrimitiveValue\nodepart[align=left]{second}val:int};

    \draw (-3,2) node[class,fill=gray!30] (ret) {ControlState};

    \path (sf.west)+(0,-0.3) coordinate (sfh);
    \draw[->] (s) |- node[below,pos=0.7]{active frame}(sfh);

    \draw[->,dotted,thick] (sf)  -- ++(-3.7,0)node[above,pos=0.4]{return state}  -- ++(0,-1.7) -- ++(-0.5,0) -| (ret);

    \path (sf.east)+(0,0.3) coordinate (sfh);
    \draw[->] (sf.north)++(0.2,0) -- ++(0,0.3) -- node[above](xx){next} ++(0.9,0) |- (sfh);

    \draw (sf) edge[diamond-] (var);
    \draw (var) edge[open diamond-] (val);

    \draw (sf) edge[->] node[left]{current object} (obj);

    \draw (s) edge[diamond-] (obj);

    \draw (refval) edge[->] node[below]{refers to} (obj);

    \draw[diamond-] (obj.east)++(0,0.25) -| node[pos=0.3,above]{attribute} (var);

    \path ($(refval.north)!.5!(primval.north)$) ++(0,.6) coordinate (h);

    \draw (refval) |- (h) -| (primval);
    \draw (h) edge[-open triangle 60] (val);

    \draw (s)++(-3,0)node{Control} edge[o-] (s);

    \draw[] node[draw,fit=(s)(sf)(val)(primval)(xx),inner sep=10pt] (l){};
    \draw (l.north) node[anchor=north](x) {\guillemotleft Storage\guillemotright};
    \addcompolabel{l}

  \end{tikzpicture}

  \caption{Simplified type graph of a storage model with stack-based recursion and an object heap}
  \label{fig:mm_storage}
\end{figure}

\myparagraph{Example: Storage Model.}
The different (sub)components of our \confG can be defined and typed according to the needs of different language semantics. In all of the execution models of \scoop, for example, we require that the storage associated with handlers consists of both an object heap and a stack frame. Figure~\ref{fig:mm_storage} depicts a \emph{type graph} for this requirement, prescribing the structure of the storage components. This (simplified) example covers recursion via a linked list of stack frames, containing variables with values that are either primitive or references to objects. These references can point to objects under the control of the same handler, but may also point to objects under the control of other (possibly distributed) handlers. The stack frame maintains a pointer to the current object with respect to the handler's execution, and may also refer to a return state (via the handler's connection to control flow information) for modelling recursion. Note that the last-in-first-out nature of the stack is modelled via the corresponding semantic rules in the \ruleE.

The example does not cover all intricacies of the storage models used in Section~\ref{sec:fac:tool}. Nevertheless, the level of detail supplied by a concrete implementation of such a component is visible. Different implementations of the metamodel's components lead to different subgraphs within the \confGs, with each choice of subgraph conforming to its own type graph. The manipulation of the elements of these subgraphs is then controlled by a dedicated set of rules in the \ruleE.

\myparagraph{Rule Engine.}
The \ruleE controls the execution of semantic graph transformation rules, and thus implements a step-wise operational semantics. For each control state of a handler (encoded as a pointer from the handler to the control flow graph in the \confG), the \ruleE nondeterministically applies one of the fireable transitions permitted by the control flow graph and any guards. Upon firing a transition, side effects may occur (e.g.~changes to the handler's storage), and the handler moves to the next control state. Note that while we assume an interleaving semantics for simplicity of presentation, the parallel execution of truly independent transitions in one global semantic step would be possible in, e.g.~the \scoop implementation.

Additionally, the \ruleE encodes a number of rules that operate in the background, performing garbage collection, object initialisation, queue management, and also moving requests from handler-to-handler via the given topology abstraction. These background rules are essential to ensure progress of the overall system. For example, a synchronous call will block a handler indefinitely if the issued request is not moved across the topology abstraction and enqueued at the other end.

\myparagraph{Semantics Parameterisation.}
Similar to the compositionality of \confGs, the rules in the \ruleE are decomposed into sets that define semantics for the different (sub)components of the metamodel. For example, in the context of \scoop, we can use a set of rules defining the queueing semantics of \req or those of \qoq. In replacing a set of rules, however, the type graph of the affected component must also be replaced to ensure that the rules operate on subgraphs exhibiting the expected structure. \req, for example, expects simple request queues, but \qoq expects to operate on nested queueing structures.

Beyond a parametric treatment of request queues, we could also plug in different topology abstractions (e.g.~introducing a two-level network hierarchy to model the distribution of handlers across nodes), or investigate different storage models for the handlers (e.g.~replacing the object heap with a simple counter variable). Each plug-in simply requires a type graph for the expected structure of the subgraph, and a new set of rules for modelling the appropriate manipulations that take place on them. All plug-ins must also ensure a consistent treatment of connections to other components, ensuring that the semantics remain modular and compositional.

\myparagraph{Metamodel for the \scoop Family.}
Our metamodel is sufficiently rich to cover the main features of the \scoop family of semantics:%
\begin{inparaenum}[(i)]
  \item message-passing based concurrency;
  \item objects that are accessible only via their assigned unique handler;
  \item asynchronous and synchronous calls to separate objects;
  \item different implementations of the synchronisation components for handlers (i.e.~\req and \qoq); and
\item different distribution topologies by plugging in different topology abstractions (i.e.~for \dscoop).
\end{inparaenum}
Concrete \gts models for the \scoop family of semantics and their implementation in our workbench will be presented in Section~\ref{sec:fac:tool}.

\myparagraph{Models Beyond \scoop.}
Due to the modularity of our metamodel, it should also be relatively straightforward to cover other asynchronous, actor-like, distributed object-oriented computation frameworks, or distributed message-passing based models.

From a theoretical point of view, we can also encode distributed finite-automata models with messages exchanged over reliable unbounded \fifo queues with local infinite recursion and an unbounded number of dynamically generated automata/handlers. While strictly weaker than (i.e.~can be simulated by) our \scoop model, they would be more manageable for deriving decidability results in the context of our workbench.

A further candidate for our semantics comparison workbench is Erlang~\cite{Armstrong-VW96a}: like \scoop, it has an intricate formal semantics, with programs behaving differently depending on whether they run in a (local) multi-core or distributed context. Furthermore, it has a highly optimised underlying runtime (see e.g.~\cite{Svensson:2010}) that has been under constant development over the last two decades, and thus may be prone to introducing unexpected behaviours for older programs. Analysing Erlang programs for concurrency bugs is extremely cumbersome: while in theory a data race free actor-based language, the runtime's scheduler and global data dictionary (\ie a kind of table-based memory heavily used in Erlang software) introduce various possibilities for race conditions in real world Erlang programs~\cite{Christakis:2010}. Thus, understanding Erlang's semantics is a crucial task for programmers, language designers, and runtime developers---\ie the target groups of our semantics workbench approach. Contrary to \scoop, Erlang focuses an asynchronous message-passing concurrency. In the language of our metamodel: handlers would synchronise only via asynchronous messages over a (possibly) distributed topology of \textsc{Fifo} queues, and while the handlers would only have a relatively flat storage model, they would have more intricate ways of accessing their \textsc{Fifo} mailboxes. Also in contrast to \scoop, Erlang is a multi-paradigm language including functional features and pattern-matching, thus the translation of the original program code to the control-flow information needed in the graph-based model is not as straightforward. Extending the semantics workbench towards Erlang would be interesting future work.


\section{Formal Model and Toolchain}
\label{sec:fac:tool}

In this section we present \scoopgts, our formal \gts model for the \scoop family of semantics, which instantiates and conforms to the semantics metamodel presented in Section~\ref{sec:fac:genericmodel}. Furthermore, we describe its implementation in the \groove model checking tool for \gts, and present a wrapper that helps to automate the analysis of \scoop source code with respect to different semantics in \groove.

A companion website~\cite{companion_website} provides additional information and explanations about the formal model that were omitted from this paper due to space constraints. Furthermore, the model and wrapper are both available to download~\cite{repo}.

\subsection{Overview}

The standalone tool at the core of our toolchain consists of the \scoopgts formalisation in \groove, and a wrapper around it which allows different execution models (\req, \qoq, \dscoop) to be selected. We furthermore provide a simple compiler to generate initial configuration graphs from \scoop source code, which the different semantics can be applied to.

An overview of the toolchain is depicted in Figure~\ref{fig:tool-overview}. The different components of the toolchain are summarised below, and presented in more detail in the rest of this section.

\begin{description}
  \item[\scoopgts.] The formalisation and implementation of the different \scoop execution models
    (\qoq, \req, \dscoop). In essence, \scoopgts is a \gts
    consisting of sets of transformation rules that define the semantic components of the metamodel, and
    control programs that dictate the order of the rule applications.
    Furthermore, the associated type graphs ensure that all graphs in the system conform to a
    certain structure, which is particularly useful during development. The
    \gts is implemented in \groove, which allows us to compare properties of state spaces under different execution models.
    The implementation is presented in detail in
    Section~\ref{sec:toolchain:scoopgts}.

  \item[\groove Wrapper.] Built on top of \groove, our wrapper provides
    command-line switches for conveniently plugging in different \scoop
    semantics, processing generated graphs, and reporting feedback. This also
    includes a regression test suite that helps to maintain correctness when
    extending \scoop-\gts.

  \item[\scoopgraphs.] Instances of configuration graphs for \scoop programs, consisting of encoded control-flow graphs corresponding to the original source code, and a snapshot of the current state of handlers. When provided as an initial state, should either store the necessary handlers pre-initialised, or a root procedure allowing \scoopgts to initialise them itself.

  \item[Graph Compiler.] In order to provide a fully automatic toolchain from
    source code to analysis results, we implemented a simple compiler
    covering basic features of the \scoop language. This compiler is
    implemented in Eiffel, which allows us to use the EiffelStudio compiler
    for parsing input programs. The compiler generates \scoopgraphs that can
    then be used directly in the standalone tool.
\end{description}

\subsection{Running Example: Dining Philosophers}

\newcommand{\eselsohr}[1]{%
    \draw[white] (#1) -- +(-0.2,0) -- +(0,-0.2) -- cycle;
    \draw[black] (#1)  +(-0.2,0) -- +(-0.2,-0.2) -- +(0,-0.2) -- cycle;
}

\begin{figure}
  \centering

  \begin{adjustbox}{max width=\textwidth,max height=\textheight}
  \begin{tikzpicture}
    \draw[fill=red!20] (5.8,1.7) coordinate (x);
    \draw[fill=red!30] (x)++(-0.2,-0.3) coordinate (y) -- ++(5,0) -- ++(0,-4.5) coordinate (l2) -- ++(-5,0) -- (y);
    \draw ($(y)!.5!(l2)$) coordinate (z) (z|-l2) node[anchor=south]{Standalone Tool};

    \draw[fill=blue!30] (-6,1) rectangle ++(8,-3);
    \draw (-2,-2) node[anchor=south]{EiffelStudio};

    \begin{scope}[every node/.style={fill=white}]
      \draw (-5,0) node[rectangle,minimum height=1cm,draw] (scoop) {\tt.scoop};
      \draw (-3,0) node[rectangle,draw,rounded corners=6pt] (flatten) {flatten};
      \eselsohr{scoop.north east}
      \draw (-1,0) node[draw,rectangle,rounded corners=6pt] (genmod) {translate};
      \draw (2,0) node[rectangle,minimum height=1cm,draw] (sg) {\scoopgraph};
      \eselsohr{sg.north east}
      \draw (5.6,0) node[rectangle,minimum height=1cm,draw] (gxl) {\tt.gxl};
      \eselsohr{gxl.north east}
      \draw (9,-1) node[draw,rectangle,minimum height=3cm,minimum width=2cm,rounded corners=6pt] (wrapper) {};
      \draw (wrapper) node[draw,rectangle,rounded corners=6pt,minimum height=1cm] (groove) {\groove};
      \draw (wrapper.north)+(0,-0.1) node[anchor=north] {wrapper};
      \draw (7,1) node[draw,rectangle] (params) {\scoopgts};
      \draw (5.6,-2) node[rectangle,minimum height=1cm,draw] (result) {\tt.txt};
      \eselsohr{result.north east}
    \end{scope}

      \draw[double,thick,-implies] (scoop) -- (flatten);
      \draw[double,thick,-implies] (flatten) -- (genmod);
      \draw[double,thick,-implies] (genmod) -- (sg);
      \draw[double,thick,-implies] (sg) -- node[above]{\small input graph} (gxl);
      \draw[double,thick,-implies] (gxl)--  (gxl-|wrapper.west);
      \draw[double,thick,-implies] (result-|wrapper.west) -- node[above]{\small result}(result);
%
      \draw[double,thick,-implies] (params) -| (wrapper);

  \end{tikzpicture}
  \end{adjustbox}
  \caption{Overview of the \scoopgts toolchain}
  \label{fig:tool-overview}
\end{figure}
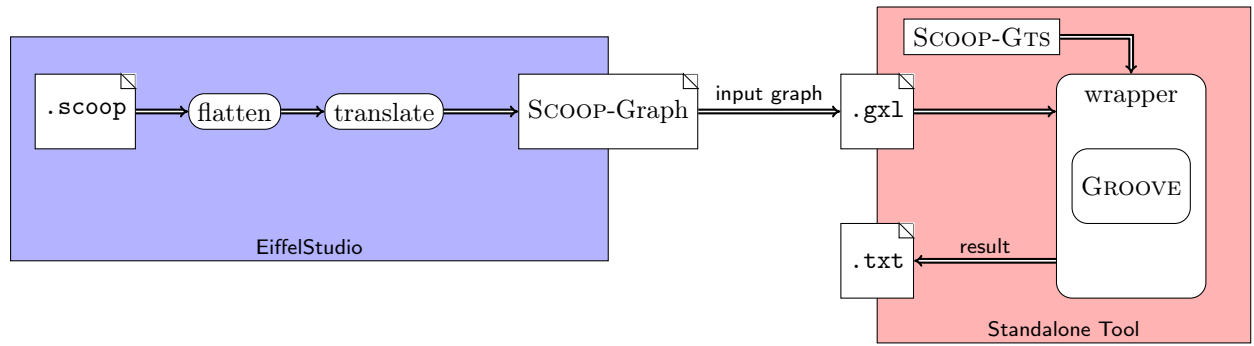

To illustrate the different parts of the formal model and toolchain, and to demonstrate a semantic inconsistency, we will use an expanded version of the dining philosophers program we introduced earlier (Section~\ref{sec:fac:scoop}).
Here, we provide only the most important excerpts of the \scoop code, adapted from an official example provided with the \scoop release~\cite{SCOOP-EiffelStudio-Reference}. The full program is available to download from our toolchain repository~\cite{repo}, and further explanations are available on our companion website~\cite{companion_website}.

Listing~\ref{listing:dp:make} contains the main creation method \incode{make} of the program, which is responsible for creating (i.e.~initialising) the forks, as well as the philosophers that will compete for them. Note that both forks and philosophers are created as objects of \incode{separate} type, meaning that all of them have their own, distinct handlers. Each philosopher object points to two fork objects with distinct handlers, and each fork object is pointed to by two philosophers, with the cyclic structure ensured by the left fork of the first philosopher being assigned as the right fork of the final philosopher. Upon the creation of each philosopher, note that its concurrent behaviour is triggered by the separate block \incode{launch_philosopher}, which asynchronously requests the philosopher's \incode{live} method.

We remark that in \dscoop, the programmer would set up different nodes (each running a \dscoop instance) manually, before reaching this step. Creating distributed objects then boils down to sending requests across the network to existing remote handlers that have creation methods available to them. In our model of \dscoop, we re-interpret this manual intialisation step, instead treating the creation of \emph{any} separate object as the creation of a new node with that new object and handler, i.e.~a scenario in which a program is as distributed as possible. In the case of \incode{make}, all the forks and philosophers would be created on their own nodes and communicate across the network.

\begin{eiffelcode}[caption={\texttt{make} method for initialising philosophers and forks},label=listing:dp:make]
make
    -- Create philosophers and forks
    -- and initiate the dinner.
  local
    first_fork, left_fork, right_fork: separate FORK
    a_philosopher: separate PHILOSOPHER
  do
    from
      i := 1
      create first_fork.make
      left_fork := first_fork
    until
      i > philosopher_count
    loop
      if i < philosopher_count then
        create right_fork.make
      else
        right_fork := first_fork
      end
      create a_philosopher.make (i, left_fork, right_fork, round_count)
      launch_philosopher (a_philosopher)
      left_fork := right_fork
      i := i + 1
    end
  end

  launch_philosopher (philosopher: separate PHILOSOPHER)
      -- Launch a_philosopher.
    do
      philosopher.live
    end
\end{eiffelcode}

Listing~\ref{listing:philosopher} contains the \incode{live} method of the \incode{PHILOSOPHER} class, which repeatedly calls a method that is supposed to simulate the philosopher exclusively holding its forks. There exist a number of different ways in which we could try to implement eating. All of them involve separate blocks over forks, but differ over whether the forks are controlled at the same time (``eagerly'') or in sequence (``lazily''), and whether the bodies of the separate blocks request any methods on those forks. Several implementations are provided: (i)~\incode{eat\_no\_statements}, which picks up forks eagerly but does not issue requests in the method body; (ii)~\incode{eat}, which is eager and asynchronously issues requests on forks; and (iii)~\incode{bad\_eat}, which picks up the forks lazily and issues asynchronous commands once they are obtained.

\begin{eiffelcode}[caption={\texttt{PHILOSOPHER} code},label=listing:philosopher]
live
  do
    from
    until
      times_to_eat < 1
    loop
      -- Philosopher `Current.id' waiting for forks.
      eat (left_fork, right_fork)
      --bad_eat
      -- Philosopher `Current.id' has eaten.
      times_to_eat := times_to_eat - 1
    end
  end

eat_no_statements (left, right: separate FORK)
    -- Eat
  do
  end

eat (left, right: separate FORK)
    -- Eat, having acquired `left' and `right' forks.
  do
    left.use
    right.use
  end

bad_eat
    -- Eat by first getting access to the `left' fork,
    -- then the `right' one.
  do
    pickup_left_then_right (left_fork)
  end

pickup_left_then_right (left: separate FORK)
  do
    pickup_right_and_eat (left, right_fork)
  end

pickup_right_and_eat (left, right: separate FORK)
  do
    left.use
    right.use
  end
\end{eiffelcode}

Under \req, both \incode{eat} and \incode{eat_no_statements} reflect valid solutions to the dining philosophers program, with the former being used in an officially provided example program~\cite{SCOOP-EiffelStudio-Reference}. Yet these are examples of developers mixing the programming abstractions of \scoop with the details of a particular execution model: in their mental models, forks had become synonymous with locks, and the process of entering the separate blocks had become synonymous with simultaneously acquiring them (or specifically, the locks on their request queues). Thus, no two adjacent philosophers can be in their separate blocks at the same time. With the \qoq semantics, however, forks cannot be viewed as locks in this way unless the separate blocks synchronise on them both (i.e.~with queries). Since they do not, the programs no longer represent correct solutions to the dining philosophers problem under \qoq, despite still respecting the high-level reasoning guarantees of the \scoop abstractions. Analogously, acquiring forks in turn can cause \incode{bad_eat} to deadlock under \req, but not under \qoq (since there is no blocking).

Our semantics comparison workbench aims to uncover discrepancies in idiomatic usages of \scoop's abstractions such as these. We later show how general-purpose rules can be used to reveal discrepancies such as the existence of deadlocks, and how specialised rules can be used to reveal discrepancies in program-specific properties, such as implementing dining philosophers correctly. We use combinations of these implementations (lazy and eager; with and without commands) in our evaluation, along with other typical concurrency benchmarks.

\subsection{\scoopgts}
\label{sec:toolchain:scoopgts}

At the heart of our workbench is \scoopgts, our parameterisable formal model for the \scoop family of semantics, conforming to the general metamodel of Section~\ref{sec:fac:genericmodel}. We implemented the transformation system in \groove~\cite{Ghamarian-MRZZ12a}, a state-of-the-art, general-purpose tool for \gts analyses.
The tool provides a GUI that allows us to draw graphs visually, which are then stored in the Graph eXchange Language format, an XML-like language for representing graph structures (designed to facilitate the exchange of graphs between different tools).
In addition to the general state space analyses it provides out of the box, \groove also supports the visual simulation of individual rule steps, which is invaluable for testing and validating the model.


\paragraph{\scoopgraphs.}
We refer to the configuration graphs in \scoopgts as \scoopgraphs. Recall that these represent a snapshot of the current state of the handlers. In the context of \scoop and \dscoop, the control flow information subgraph encodes both methods and classes; the handler subgraphs encode object heaps, stack frames, and some queuing structure (\req or \qoq); and the topology abstraction subgraph consists of (separate) object references, and (in \dscoop) information about the nodes that handlers are located on.

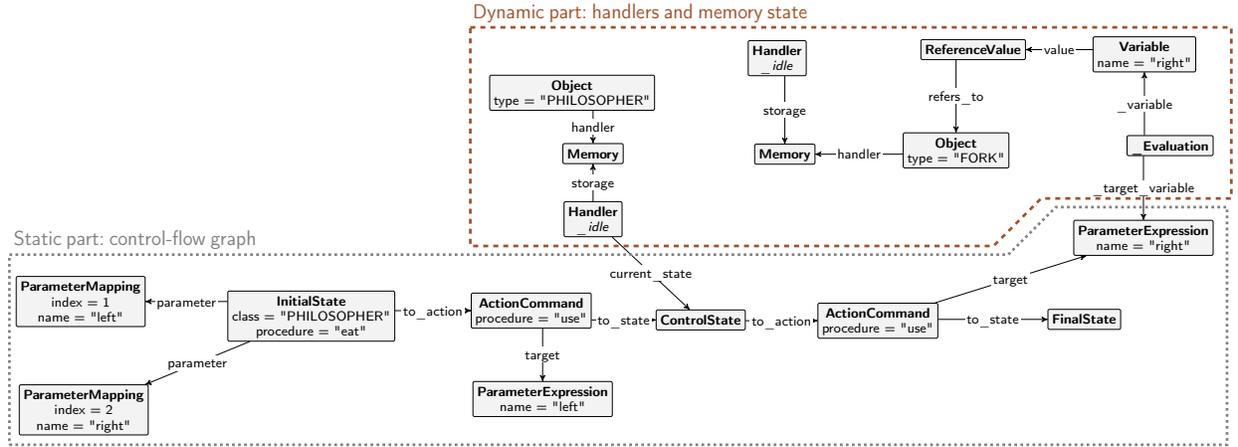
\begin{figure}
  \centering
  \begin{adjustbox}{width=\textwidth,max height=\textheight}
%
\begin{tikzpicture}[scale=\tikzscale]
\node[basic_node] (n153) at (19.970, -12.820) {\ml{\textbf{ControlState}}};
\node[basic_node] (n149) at (13.452, -12.626) {\ml{\textbf{ParameterMapping}\\index = 1\\name = "left"}};
\node[basic_node] (n154) at (21.832, -12.828) {\ml{\textbf{ActionCommand}\\procedure = "use"}};
\node[basic_node] (n150) at (13.484, -13.766) {\ml{\textbf{ParameterMapping}\\index = 2\\name = "right"}};
\node[basic_node] (n151) at (18.188, -12.722) {\ml{\textbf{ActionCommand}\\procedure = "use"}};
\node[basic_node] (n155) at (24.624, -11.956) {\ml{\textbf{ParameterExpression}\\name = "right"}};
\node[basic_node] (n152) at (18.308, -13.652) {\ml{\textbf{ParameterExpression}\\name = "left"}};
\node[basic_node] (n156) at (24.002, -12.814) {\ml{\textbf{FinalState}}};
\node[basic_node] (n148) at (15.876, -12.780) {\ml{\textbf{InitialState}\\class = "PHILOSOPHER"\\procedure = "eat"}};
\node[basic_node] (n327) at (22.654, -11.036) {\ml{\textbf{Object}\\type = "FORK"}};
\node[basic_node] (n329) at (22.834, -9.976) {\ml{\textbf{ReferenceValue}}};
\node[basic_node] (n340) at (18.618, -10.430) {\ml{\textbf{Object}\\type = "PHILOSOPHER"}};
\node[basic_node] (n345) at (18.838, -11.070) {\ml{\textbf{Memory}}};
\node[basic_node] (n347) at (18.838, -11.760) {\ml{\textbf{Handler}\\\textit{\_idle}}};
\node[basic_node] (n471) at (24.634, -10.026) {\ml{\textbf{Variable}\\name = "right"}};
\node[basic_node] (n491) at (24.904, -10.986) {\ml{\textbf{\_Evaluation}}};
\node[basic_node] (n334) at (20.774, -10.066) {\ml{\textbf{Handler}\\\textit{\_idle}}};
\node[basic_node] (n332) at (20.854, -11.076) {\ml{\textbf{Memory}}};

\path[basic_edge](n148.west |- 13.452, -12.626) -- node[lab] {\ml{parameter}} (n149) ;
\path[basic_edge] (n154)  -- node[lab] {\ml{target}} (n155) ;
\path[basic_edge](n151.south -| 18.308, -13.652) -- node[lab] {\ml{target}} (n152) ;
\path[basic_edge] (n148)  -- node[lab] {\ml{parameter}} (n150) ;
\path[basic_edge](n153.east |- 21.832, -12.828) -- node[lab] {\ml{to\_action}} (n154) ;
\path[basic_edge](n148.east |- 18.188, -12.722) -- node[lab] {\ml{to\_action}} (n151) ;
\path[basic_edge](n151.east |- 19.970, -12.820) -- node[lab] {\ml{to\_state}} (n153) ;
\path[basic_edge](n471.west |- 22.834, -9.976) -- node[lab] {\ml{value}} (n329) ;
\path[basic_edge](n329.south -| 22.654, -11.036) -- node[lab] {\ml{refers\_to}} (n327) ;
\path[basic_edge] (n347)  -- node[lab] {\ml{current\_state}} (n153) ;
\path[basic_edge](n491.north -| 24.634, -10.026) -- node[lab] {\ml{\_variable}} (n471) ;
\path[basic_edge](n347.north -| 18.838, -11.070) -- node[lab] {\ml{storage}} (n345) ;
\path[basic_edge](n154.east |- 24.002, -12.814) -- node[lab] {\ml{to\_state}} (n156) ;
\path[basic_edge](n327.west |- 20.854, -11.076) -- node[lab] {\ml{handler}} (n332) ;
\path[basic_edge](n334.south -| 20.854, -11.076) -- node[lab] {\ml{storage}} (n332) ;
\path[basic_edge](n340.south -| 18.838, -11.070) -- node[lab] {\ml{handler}} (n345) ;
\path[basic_edge](n491.south -| 24.624, -11.956) -- node[lab] {\ml{\_target\_variable}} (n155) ;

\path (n150.south west)++(-.1, -.1) coordinate (cfgleft);
\path (n150.south west)++(-.1, 1.9) coordinate (cfgleftup);
\path (n156.north east)++(0.1, 0.4) coordinate (cfgright);

\path (n347.south west)++(-1, -0.3) coordinate (dynleft);
\path (n340.north west)++(-0.2, 0.5) coordinate (dynleftup);
\path (n471.north east)++(0.5, 0.1) coordinate (dynright);

\draw (cfgleft)++(0,2.3) node[anchor=north west, text width=6cm, gray] {{\large Static part: control-flow graph}};
\draw (dynleft)++(0, 2.8) node[Sienna,anchor=north west, text width=8cm] {{\large Dynamic part: handlers and memory state}};
\draw[ultra thick, Sienna, dashed] (dynleftup) -- ++(8, 0) -- ++(0, -1.8) -- ++(-2, 0) -- ++(-0.5, -0.5) -- ++(-5.5, 0) -- (dynleftup);
\draw[ultra thick, gray, dotted] (cfgleft) -- ++(12.8, 0) -- ++(0, 2.5) -- ++(-1.8, 0) -- ++(-0.5,-0.5) -- (cfgleftup) -- (cfgleft);

\end{tikzpicture}
  \end{adjustbox}
   \caption{Control-flow graph for the \texttt{eat} method and a handler executing it. The parameter of the \texttt{use~(right)} command has been evaluated and refers to a fork on a different handler. The static part remains unchanged throughout the simulation, but the handler in the dynamic part moves along the control-flow graph}
  \label{fig:cfg}
\end{figure}

Part of a \scoopgraph is shown in Figure~\ref{fig:cfg}, which depicts the
simple control-flow graph of the \incode{eat} method in the lower half, and
the state of a handler in the upper half. The control-flow graph itself is
static: there are no rules that directly modify these nodes and edges.
However, handlers can fire transitions encoded in these control-flow
graphs. We model this using an edge labelled \texttt{current\_state} for each
handler that is currently executing a method. As handlers execute actions,
they move along the control-flow graphs in the expected way. Note that \dscoop control-flow graphs do not contain additional information specific to distributed computing. This is because in terms of the abstractions, programmers can use remote \incode{separate} objects in the same way as those residing on the same node; the only difference is in the topology abstraction.

{Note that initial \scoopgraphs, such as those generated by our simple compiler, contain only the static control-flow part. There are no edges or nodes related to a particular runtime semantics, and as a result, we can use the same initial graphs for a given program and decide later on the semantics we want to use to simulate the program.}

\paragraph{Control Programs.}\label{sec:controlprograms}
Our current rule engine includes around 120 transformation rules covering
local computations, execution of commands and queries, runtime management,
queueing, and other activities. In most situations, we do not want to explore all
possible rule applications. Instead, we often have situations where we only
want to apply a certain (set of) rules. For example, when a handler finishes
executing a method, it may be possible that there are leftover nodes and edges
in the graph. To clean this up, we provide \emph{bookkeeping} (or \emph{background}) rules. Of
course, it makes sense to apply these as soon as possible instead of allowing
other rules to be applied (e.g. rules that advance the execution of other
handlers) since their effects are local to a handler and independent of all others. One way to ensure a certain order of execution is to design rules
such that they can only be applied when we want them to. Here, negative
application conditions are commonly used. For example, we could insert a node
\texttt{CleanupInProgress} when a method execution is finished and place it
in all cleanup rules (as a node that needs to be present in order for the rule to match).
Similarly, we would add a negative application condition
for this node in all other rules (making sure that the rule only gets applied if the graph is \textit{not} in a cleanup state).
However, we use this strategy only sparingly in our
implementation, since the resulting rules tend to become large and contain
many seemingly unrelated nodes.

Fortunately, \groove provides another way
to restrict rule application orders, namely \emph{control programs}. These
programs allow us to specify execution orders and avoid having cluttered rules. The overall effect is that the firing of a transition in the control-flow graph appears to happen \emph{atomically}---regardless of the number of rules involved---meaning that we exclude unnecessary interleavings on local bookkeeping rules. Listing~\ref{listing:control-program} shows a simplified version of the main control program that drives the execution of \scoopgts.

\begin{lstlisting}[float=tp,style=controlprgstyle,label=listing:control-program,caption={Simplified control program (in \groove syntax) from the \scoopgts rule engine}]
initialize_model;                              // call gts rule for initialisation
while (progress & no_error) {
  for each handler p:                          // choose handlers under some scheduling strategy
    alap handler_local_execution_step(p)+;     // each handler executes local actions as long as possible
  try synchronisation_step;                    // then try (one) possible global synchronisation step
}
recipe handler_local_execution_step (p){
  try separate_object_creation(p)+;            // try local actions that are possibly applicable
  else try assignment_to_variable(p)+;
  else try ... ;                               // sequentially try all other possible actions
  try garbage_collection()+;                   // do some "garbage collection" to keep the model small
}
recipe synchronisation_step(){
  reserve_handlers | dequeue_task | ...;       // nondeterministically try to synchronise
}
...                                            // remaining recipes (core functionality)
// ---------- plug in -------------------------------------------------------------------------------
recipe separate_object_creation(p){            // provide different implementations for RQ and QoQ
  ...                                          // and parameterise the control program
}
...                                            // remaining recipes that are plugged in
\end{lstlisting}

Using these control programs furthermore allows us to perform optimisations, and
force particular rule applications when exploring other execution paths would not reveal any additional behaviours. One example occurs when multiple handlers execute local
computations (i.e.~computations that do not involve separate or remote objects). While all interleavings are possible in the actual runtimes,
we do not need the overhead of simulating all of them, since local computations do not involve any interactions between handlers. In \scoopgts, we therefore mark one handler
as \emph{active} and advance it as long as possible, until it terminates or a
statement involving separate objects is reached. Then, we activate the next
handler (in an ordered list of all handlers) and do the same until no
handlers are able to execute local computations anymore. At this point,
handlers are either idle (i.e. they have finished their execution), or they
are at a synchronisation point where other handlers are involved in the next
operation. Here, rules involving separate entities are applied in a
nondeterministic manner. This way, we ensure that all interleavings that are
of interest to us (i.e. interleavings that result in different orders and
configurations of the queues) are explored. Thanks to these optimisations, we made it feasible to perform
full state-space exploration for small programs like the running example and the ones discussed in
Section~\ref{sec:fac:analysis}.

\paragraph{Transformation Rules.}
One advantage of a graph representation of is that the graphs
can be easier to read (for smaller instances, at least). This is in particularly true for the
transformation rules in our system, since they are usually small and perform a
simple task. In an earlier prototype without control programs, the rules often
contained helper nodes and negative application conditions to make sure a rule
is only applied when appropriate. However, in our current implementation we
can avoid most of these helper elements and end up with clean, more direct rules.
Furthermore, transformation rules are expressive and atomic, with \groove able to support the matching of arbitrary-length paths and quantify over substructures.

\begin{figure}
  \centering
  \begin{adjustbox}{width=\textwidth,max height=\textheight}
    \includegraphics[width=\textwidth]{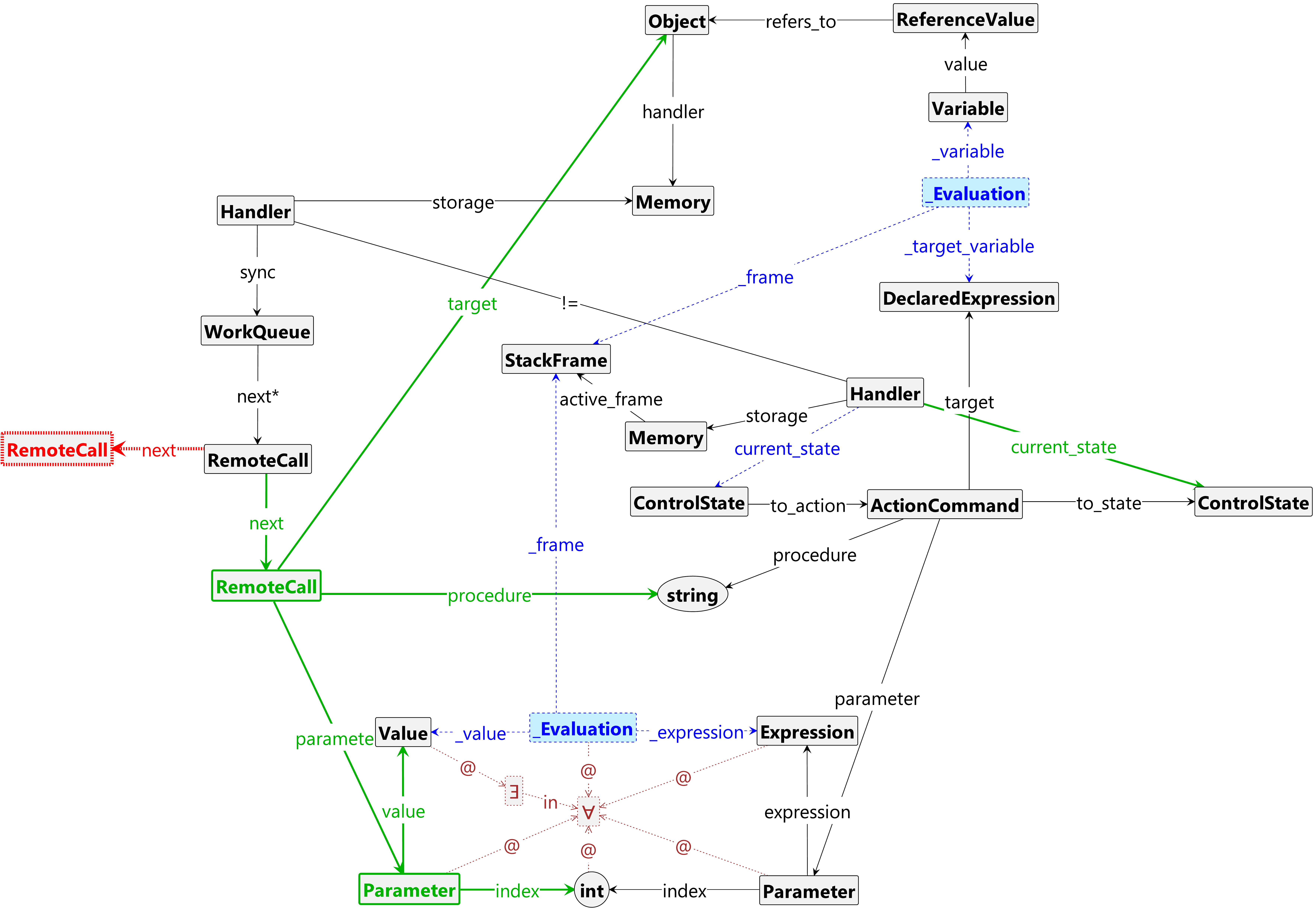}
  \end{adjustbox}
  \caption{Full, unmodified rule for enqueueing a new remote call in the \req
    semantics. The \incode{Handler} node is advanced from one
    \incode{ControlState} to another. The new remote call is appended to the
    \incode{WorkQueue} of the target handler. The tail is matched using a
    \incode{next*} edge from the work queue to the black \incode{RemoteCall}. We ensure that the tail node is matched using a negative
    application condition. Finally, the green \incode{RemoteCall} is inserted
    at the end. Note that in the lower part, we attach the evaluated
    parameters to the remote call}
  \label{fig:rule_rq_enqueue}
\end{figure}

Figure~\ref{fig:rule_rq_enqueue} shows the rule that, when applied, enqueues a request (the green \incode{RemoteCall} node) into a target's request queue---a similar task to that from our introduction to \gts in Section~\ref{sec:primer_gts} (but without the anchor node at the tail of the queue). In essence, this rule updates a handler's \incode{current\_state} by moving it across an \incode{ActionCommand} node, which represents either executing a command directly (if the target is handled by the same handler) or issuing an asynchronous request on the target handler. The rule is for the latter case: it matches a different handler (indicated by the \incode{!=} edge) when looking up the target, which is found by following the \incode{target} edge from \incode{ActionCommand} up to the evaluated value and its handler. Since the target is handled by a different handler, the request queue (\incode{WorkQueue}) of the target's handler is matched and a new \incode{RemoteCall} node is appended to the queue. In the lower part of the rule, parameters are passed by matching all indexed parameters corresponding to the method call and adding parameter nodes to a remote call. These parameter nodes then point directly to the evaluated values. Finally, the \incode{\_Evaluation} nodes are removed, as they are no longer needed once the handler has passed the action node in the control graph and the request has been issued.


\begin{figure}
  \centering
  \begin{adjustbox}{width=\textwidth,max height=\textheight}
    \includegraphics[width=\textwidth]{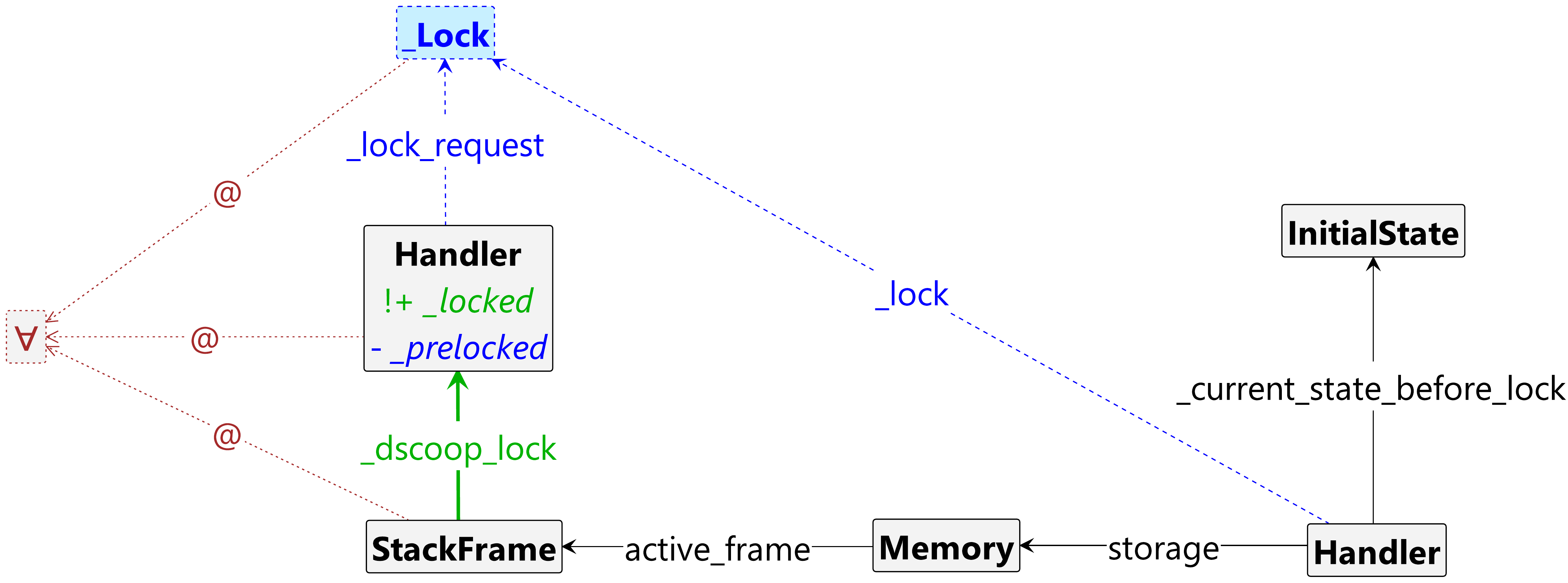}
  \end{adjustbox}
  \caption{Simplified \dscoop rule that is applied when (i)~all target
    handlers (left side) are prelocked by the executing handler (lower right),
    and (ii) the executing handler has sent lock requests to all of them.
    The \incode{\_Lock} nodes get removed upon application}
  \label{fig:rule_dscoop_lock_ack}
\end{figure}

A second example, shown in Figure~\ref{fig:rule_dscoop_lock_ack}, shows one of the few additional rules required for implementing \dscoop on top of the \qoq semantics. Once all prelocks have been obtained and lock requests sent (indicated by the \incode{\_Lock} node and its edges), this rule performs the acknowledgement of the requests on all target handlers by setting the flag \incode{\_locked}. Note that the main handler is connected to an initial state via a \incode{\_current\_state\_before\_lock} edge. Since this edge label is only used in \dscoop related rules, this means that the normal \qoq rules cannot fire and simulate past the initial state. Instead, \dscoop rules will match and the whole locking process is simulated. Once this is done (i.e.~all locks are obtained and the handler is at a state where no more \dscoop related operations need to be performed), the last \dscoop rule replaces the \incode{\_current\_state\_before\_lock} edge with a ``normal'' \incode{current\_state} edge (as used in \qoq). From this point on, the method execution will be simulated just like a \qoq program.

\paragraph{Modularity of \scoopgts.}

An important part of our formalisation is modularity. In the case of \scoop,
we have three different execution models that share certain properties. For example,
local computations, in which no synchronisation is involved, do not behave differently. Instead, we can use the same rules for all semantics. We can
achieve this using the previously discussed control programs. We split the
programs into a generic root program that covers rules that are used in all
runtimes. From these, however, several so-called \emph{recipes} (intuitively: functions over rules) are referenced that differ from runtime to runtime (most notably \texttt{synchronisation\_step()}, which nondeterministically performs one of
the next possible asynchronous tasks, and \texttt{garbage\_collection()}, which
cleans up the graph). Consequently, we have
three additional files, each one covering a specific runtime and implementing
the same recipes, but using different rules. This way, we can switch runtimes
simply by stating which control programs are enabled. Note that it is not
necessary to state which rules are enabled at different points, since only those
referenced in an active control program are applied.

Similar to control programs, type graphs are also split into several files. By
doing this, we can enable only the relevant type graphs when working on a
specific runtime. As a result, we do not end up with malformed, hard to debug
state graphs, that, for example, mix \req and \qoq semantics.

To illustrate how close two of the semantics are, we compare the implementations of
\qoq and \dscoop. With \dscoop being an extension on top of \qoq, we were able
to start by reusing the \qoq implementation. Implementing the basic
prelock/lock mechanism took no more than 4 rules (prelock request, prelock
acknowledgement, lock request, and lock acknowledgement).  We then use some
additional rules that handle method calls (in particular, entering methods
needs to toggle the new protocol before executing the standard \qoq
operations) and method returns (here, we handle releasing the locks and
cleaning up the graph). Finally, we reference these additional rules in a
control program. This program contains only the parts that differ from \qoq by
re-defining recipes that use the new rules. The vast majority of both control
programs and rules, however, stay the same between the two runtimes, making
our approach practical for extending and modifying existing implementations.

\paragraph{Errors and Consistency of Semantics.}

With the workbench, we can now check programs and runtimes for errors and
inconsistencies. One application is to verify certain properties of a given
\scoop program across different execution models. For example, one might be curious whether the dining
philosophers program can deadlock. To do this, we simply use the standalone
tool and perform a full state-space exploration. The tool reports the final
states of the \gts, i.e.~the states where no more rules can be applied.  In the
\gts, undesired behaviour is specified with \emph{error rules}, which capture certain properties
of a runtime program. These rules are checked between synchronisation points,
i.e. whenever the program branches due to multiple possible interleavings of
the execution. In case an error rule matches, it is applied and a node of
(sub)type \texttt{Error} is created. As a result, the control program
immediately stops further execution, resulting in a final state representing
the exact point at which the error occurred. This way, we can see whether
any errors occurred during exploration by simply iterating over all final states
and checking whether an error node is present in any of them.

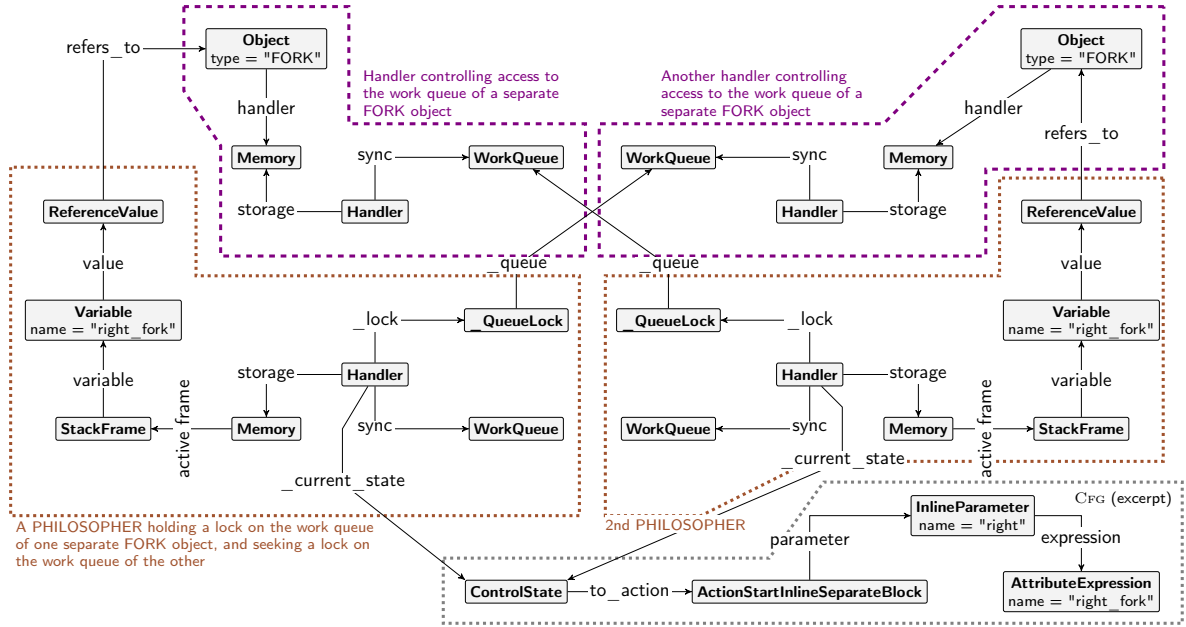
\begin{figure}
	\centering
	\begin{adjustbox}{width=\textwidth-0.8cm,max height=\textheight}
%
\begin{tikzpicture}[scale=\tikzscale]
\node[basic_node] (state) at (-0.7,0) {\ml{\textbf{ControlState}}};
\node[basic_node] (act) at (2,0) {\ml{\textbf{ActionStartInlineSeparateBlock}}};
\node[basic_node] (inparam) at (3.5,0.7) {\ml{\textbf{InlineParameter}\\name = "right"}};
\node[basic_node] (attexpr) at (4.5,-0.0) {\ml{\textbf{AttributeExpression}\\name = "right\_fork"}};

\node[basic_node] (p1) at (-2, 2) {\ml{\textbf{Handler}}};
\node[basic_node] (p2) at (2,2) {\ml{\textbf{Handler}}};
\node[basic_node] (p3) at (2,3.5) {\ml{\textbf{Handler}}};
\node[basic_node] (p4) at (-2,3.5) {\ml{\textbf{Handler}}};

\node[basic_node] (wq1) at (-0.7, 1.5) {\ml{\textbf{WorkQueue}}};
\node[basic_node] (wq2) at (0.7,1.5) {\ml{\textbf{WorkQueue}}};
\node[basic_node] (wq3) at (0.7,4) {\ml{\textbf{WorkQueue}}};
\node[basic_node] (wq4) at (-0.7,4) {\ml{\textbf{WorkQueue}}};

\node[basic_node] (ql1) at (-0.7, 2.5) {\ml{\textbf{\_QueueLock}}};
\node[basic_node] (ql2) at (0.7,2.5) {\ml{\textbf{\_QueueLock}}};

\node[basic_node] (m1) at (-3,1.5) {\ml{\textbf{Memory}}};

\node[basic_node] (m2) at (3,1.5) {\ml{\textbf{Memory}}};

\node[basic_node] (m3) at (3,4) {\ml{\textbf{Memory}}};
\node[basic_node] (m4) at (-3,4) {\ml{\textbf{Memory}}};

\node[basic_node] (sf1) at (-4.5,1.5) {\ml{\textbf{StackFrame}}};
\node[basic_node] (v1) at (-4.5,2.5) {\ml{\textbf{Variable}\\name = "right\_fork"}};
\node[basic_node] (rv1) at (-4.5,3.5) {\ml{\textbf{ReferenceValue}}};

\node[basic_node] (sf2) at (4.5,1.5) {\ml{\textbf{StackFrame}}};
\node[basic_node] (v2) at (4.5,2.5) {\ml{\textbf{Variable}\\name = "right\_fork"}};
\node[basic_node] (rv2) at (4.5,3.5) {\ml{\textbf{ReferenceValue}}};

\node[basic_node] (o1) at (-3,5) {\ml{\textbf{Object}\\type = "FORK"}};
\node[basic_node] (o2) at (4.5,5) {\ml{\textbf{Object}\\type = "FORK"}};

\begin{scope}[every edge/.style={basic_edge,->},every node/.style={lab,font=\sf}]
  \draw[basic_edge] (p1) -| node{storage} (m1);
  \draw[basic_edge] (p2) -| node{storage} (m2);
  \draw[basic_edge] (p3) -| node{storage} (m3);
  \draw[basic_edge] (p4) -| node{storage} (m4);

  \draw[basic_edge] (m1) -- node[above,rotate=90] {active frame} (sf1);
  \draw[basic_edge] (m2) -- node[above,rotate=90] {active frame} (sf2);

  \draw[basic_edge] (sf1) -- node {variable} (v1);
  \draw[basic_edge] (sf2) -- node {variable} (v2);

  \draw[basic_edge] (v1) -- node {value} (rv1);
  \draw[basic_edge] (v2) -- node {value} (rv2);

  \draw[basic_edge] (rv1) |- node {refers\_to} (o1);
  \draw[basic_edge] (rv2) -- node {refers\_to} (o2);

  \draw[basic_edge] (o1) -- node {handler} (m4);
  \draw[basic_edge] (o2) -- node {handler} (m3);

  \draw[basic_edge] (p1) |- node {sync} (wq1);
  \draw[basic_edge] (p2) |- node {sync} (wq2);
  \draw[basic_edge] (p3) |- node {sync} (wq3);
  \draw[basic_edge] (p4) |- node {sync} (wq4);

  \draw[basic_edge] (p1) |- node {\_lock} (ql1);
  \draw[basic_edge] (p2) |- node {\_lock} (ql2);

  \draw[basic_edge] (ql1) -- +(0,0.5) node {\_queue} -- (wq3);
  \draw[basic_edge] (ql2) -- +(0,0.5) node {\_queue} -- (wq4);

  \draw[basic_edge] (p1) -- ++(-0.3,-0.5) -- ++(0,-0.5) node {\_current\_state} -- (state.north west);
  \draw[basic_edge] (p2) -- ++(0.3,-0.3) -- ++(0,-0.5) node {\_current\_state} -- (state.north east);

  \draw[basic_edge] (state) -- node{to\_action}(act);
  \draw[basic_edge] (act) |- node[pos=0.3]{parameter}(inparam);
  \draw[basic_edge] (inparam) -| node[pos=0.7]{expression}(attexpr);

\end{scope}




\begin{pgfonlayer}{background}
  \path (m1) +(0,-0.8) coordinate (dummy0);
  \path (rv1.north west)+(-0.3,0.3) coordinate (dummy1);
  \path (rv1.south east)+(0.3,0) coordinate (dummy2);
  \path (ql1.north east)+(0.1,0.3) coordinate (dummy4);
  \draw[ultra thick, dotted,Sienna]  (dummy1) -- (dummy1 |- dummy0) -- (dummy4 |- dummy0) -- (dummy4) -- (dummy4 -| dummy2) -- (dummy2|-dummy1) -- (dummy1);
  \draw[Sienna] (dummy1|-dummy0) node[anchor=north west,text width=6.8cm]
  {A PHILOSOPHER holding a lock on the work queue of one separate FORK object, and seeking a lock on the work queue of the other };

  \path (rv2.north east)+(0.2,0.2) coordinate (dummy0);
  \path (rv2.north west)+(-0.2,0.2) coordinate (dummy1);
  \path (ql2.north west)+(-0.1,0.3) coordinate (dummy3);
  \path (m2) +(-2,-0.8) coordinate (dummy4);
  \path (m2) +(-1,-0.3) coordinate (dummy5);

  \draw[Sienna,ultra thick, dotted] (dummy0) -- (dummy1) |- (dummy3) |- (dummy4) -- (dummy5) -| (dummy0);
  \path (dummy3|-dummy4) coordinate (label2ndphil);

  \path (o1.north west) +(-0.2,0.2) coordinate (dummy0);
  \path (o1.south west) +(-0.2,-0.2) coordinate (dummy01);
  \path (m4.north west)+(-0.1,0) coordinate (dummy1);
  \path (p4.south)+(0,-0.3) coordinate (dummy2);
  \path (wq4.north east)+(0.2,0.2) coordinate (dummy3);
  \path (o1.north east) +(0.2,0.2) coordinate (dummy4);

  \draw[dashed, ultra thick, Purple] (dummy0) -- (dummy01) -- (dummy1) -- (dummy1|- dummy2) -- (dummy3|- dummy2) -- (dummy3) -- (dummy3-|dummy4) -- (dummy4) -- (dummy0);

  \draw[Purple] (dummy3) node[anchor=south east, text width=4cm]
    {Handler controlling access to the work queue of a separate FORK object};

  \path (o2.north east)+(0.2,0.2) coordinate (dummy0);
  \path (o2.north west)+(-0.2,0.2) coordinate (dummy1);
  \path (m3.north west)+(0,0.2) coordinate (dummy12);
  \path (wq3.north west)+(-0.2,0.2) coordinate (dummy2);
  \path (p3.south)+(0,-0.3) coordinate (dummy3);
  \path (m3.south east)+(0.3,0) coordinate (dummy4);

  \draw[dashed,ultra thick,Purple] (dummy0) -- (dummy1) -- (dummy12) -- (dummy2) |- (dummy3) -| (dummy4) -| (dummy0);

  \draw[Purple] (dummy12) node[anchor=south east, text width=4cm]
    {Another handler controlling access to the work queue of a separate FORK object};


    \draw[ultra thick,dotted,gray] (state.north west)++(-0.2,0.2) -- ++(3,0) -- ++(0.5,0.7) -- ++(3.3,0)coordinate(n) -- ++(0,-1.3) -| cycle;
    \draw (n) node[anchor=north east,inner sep=5pt] {\cfg (excerpt)};

\end{pgfonlayer}
  \draw[Sienna] (label2ndphil) node[inner sep=0pt,outer sep=0pt,anchor=north west,fill=white]
  {2nd PHILOSOPHER};

\end{tikzpicture}
	\end{adjustbox}
  \caption{Reachable deadlock under \req for the lazy philosophers program (\incode{bad_eat}), simplified from the \groove output, with additional highlighting and information in colour. Both philosophers are waiting to acquire the lock on the (queue of the) fork held by the other}
	\label{fig:fac:deadlock_rq}
\end{figure}

An example of two handlers (philosophers) in a deadlock is shown in Figure~\ref{fig:fac:deadlock_rq}.
Each philosopher is trying to obtain a lock on the handler of its right fork, but of course, that
one is already locked. Since no philosopher gives up a lock, the program cannot proceed.
An error rule for deadlock matches the pattern involving the cycle between the control state, handlers executing the method, queue locks (on the handlers of the forks), and variable targets (referencing the already locked fork).

While errors related to the execution model are usually universal (e.g. deadlocks can
be expressed in similar ways in all runtimes and are independent of the
simulated program), it is also possible that certain programs have further
properties that we want to check. For example, in the dining philosophers
program with commands (\incode{eat}), none of the execution models produce a deadlock error. However,
when executed under \qoq, the program is not a valid implementation of the
dining philosophers (since philosophers sharing a fork can enqueue commands to
their private subqueues in the fork's handler at the same time). By creating a
custom error rule for this program that matches whenever two handlers are
executing the \incode{eat} method (or one of its variants), we can show that
this condition arises with the \qoq runtime, but not with the \req runtime.

\paragraph{Soundness/Faithfulness of \scoopgts.}
Due to the varying levels of detail in the previous formalisations of the execution models (and complete lack of formalisations of their corresponding implementations/runtimes), there is no universal way to formally prove \scoopgts's faithfulness to them. In the following, we describe the techniques we applied to establish confidence in its soundness despite this challenge. We remark that \scoopgts currently does not support some advanced programming mechanisms of the Eiffel language (e.g.~exceptions, agents), but could straightforwardly be extended to cover them.

We were able to conduct expert interviews with the researchers proposing the execution models and the programmers implementing the \scoop compiler and runtimes (\ie as part of EiffelStudio), which helped to improve our confidence that \scoopgts faithfully covers their behaviour. Here, \scoopgts's advantage of being a visually accessible notation was extremely beneficial, as we were able to directly use simulations in \groove during the interviews, which were understood and accepted by the interviewees. Before the interviews, we would prepare configurations (graphs) representing interesting scenarios, and would click through rule applications in \groove together with the experts. In a sense, our formal model partially mapped to how they would informally sketch the execution models for us on a whiteboard.

In addition, we compared \groove simulations of the executions of \scoop programs (those based on the benchmarks of Section~\ref{sec:fac:analysis}) against their actual execution behaviour in the official \scoop IDE and compiler (both the current release that implements \qoq, and an older one that implemented \req; for \dscoop there is no official release of a compiler/runtime yet as it currently exists only as a research prototype). Again, this augmented our confidence.

Furthermore, we were able to compare the \qoq execution model with the structural operational semantics for \qoq provided in \cite{West-NM15b}. Unfortunately, the provided semantic rules focus only on a much simplified core, preventing a rigorous bisimulation proof exploiting the algebraic characterisations of \gts. We can, however, straightforwardly implement and simulate them in our model.

Our \dscoop model is based on the \qoq model, incorporating an abstraction of the underlying network topology and an implementation of the locking protocol in our model's scheduler. Again, we compared our model to the informal (but detailed) description in~\cite{Schill-PM16a}, \eg by testing the simulation of the underlying message exchanges of the locking protocol in our model. Additionally, we interviewed the developer of \dscoop based on simulation runs of our model in \groove. Regrettably, the only existing formal model of \dscoop was not an operational one but rather a ``context-sensitive grammar for a language composed of messages on a timeline''~\cite{Schill-thesis}, thus precluding a direct formal semantic comparison.

\myparagraph{\groove Wrapper and \scoopgraph Compiler.}

\scoopgts is the main outcome of our work and can be used directly with the
\groove binaries. However, it is tedious to do so, since it requires knowledge
about the implementation of \scoopgts. To mitigate this problem, we provide
a simple wrapper utility around \groove that operates in the domain of
\scoopgts. The utility provides a command-line interface to configure and
instantiate \scoopgts. It then uses \groove to run the state-space exploration. As
opposed to \groove itself, which provides generic output, we can parse the
final states and check for the existence of \texttt{Error} nodes and other
properties of the graph. Finally, these findings are reported. In addition to
this scenario, we also use the wrapper utility for testing purposes and for
generating the benchmark results that are presented in this paper.

While not part of the current distribution, we also implemented a simple
compiler that translates \scoop programs into \scoopgraphs. This helps make all aspects of the toolchain practical, since we do not have to specify initial graphs
manually nor annotate the source code. Instead, we can use unmodified code to
generate graphs and verify concurrency properties. The compiler is implemented
in Eiffel on top of the
EiffelStudio\footnote{\url{https://www.eiffel.com/eiffelstudio/}} compiler, and is currently being integrated into a research branch of the IDE~\cite{Tschannen-FNM11a}.
As the control flow information can be mapped directly back to the source code, we can provide feedback on the analyses to the programmer based on concrete lines of code.


\section{Tool Case Studies and Evaluation}
\label{sec:fac:analysis}

In this section, we present an evaluation of our toolchain. We apply it to a selection of \scoop benchmarks consisting of small, self-contained programs that represent idiomatic usages of the language's concurrency abstractions. By simulating these
programs using \scoopgts and \groove, we show that it is feasible to
explore the full state-spaces of such a benchmark set and check the consistency of properties across different execution models.

\paragraph{Benchmark Selection.}
Our aim was to devise a set of representative programs that cover typical
usages of \scoop's concurrency mechanisms. In order to be able to explore the
full state-spaces, these programs were selected with the capabilities of the workbench in mind: even though our \gts rules and control programs limit the amount of unnecessary interleaving, larger programs will still suffer from the state-space explosion problem. We therefore based our
benchmark programs on the official, documented \scoop
examples~\cite{SCOOP-EiffelStudio-Reference} and some classical
synchronisation problems. We implemented these programs in \scoop, and used
our compiler to automatically generate the corresponding initial graphs, i.e.~encoding the control-flow of the original programs.  Everything necessary to reproduce the
benchmarks in this section is available from our online repository~\cite{repo}.

We selected the following programs: dining philosophers (as presented in
Section~\ref{sec:fac:tool}) with its two implementations for picking up forks
that exploited the implicit locking of \req (eagerly, by picking them atomically, or lazily, by picking them in sequence---\texttt{eat} and \texttt{bad\_eat} from Listing~\ref{listing:philosopher}
respectively); another two
variants of the dining philosophers without any commands in the separate blocks;
single-element producer consumer, which uses a mixture of commands, queries,
and wait conditions; and finally, barbershop and dining savages
(adapter from ``The Little Book of Semaphores''~\cite{little-book-of-semaphores}), both of which use a similar mix
of features. These programs cover different usages of \scoop's language
mechanisms and are well-understood examples in concurrent programming. Note
that while our compiler supports inheritance by flattening the used classes,
these examples do not use inheritance; in particular, no methods from the
implicitly inherited class \incode{ANY} are used. By not translating these
methods into the initial graphs, we obtain considerably smaller graphs (which
impacts the exploration speed, but not the sizes of the generated transition
systems).

Table~\ref{tab:evaluation} summarises metrics for the mentioned programs,
where the columns are reported as follows:
\begin{description}
  \item[Initial Graph.] Name of the program that is executed. These initial graphs
    are direct outputs from the compiler without further modifications. Since
    the initial graphs consist only of the control-flow graphs (i.e. the static
    part of a \scoopgraph), there are no differences between the individual
    runtimes: all of them start with the same initial graph, but once we select the
    control programs and transformation rules, the evolution of the graph
    reflects the selected runtime's behaviour.
  \item[Runtime.] Parameterised \scoop semantics: \req, \qoq, or \dscoop.
    Each semantics has its own control programs and transformation rules (with
    shared elements). The wrapper utility allows us to select the execution model that should be used via a simple command-line switch.
  \item[Configurations.] The number of proper configurations in the exploration. Note that in
    this context, a state is counted whenever a full local execution step or
    synchronisation step (cf.~Listing~\ref{listing:control-program}) is applied.
    Intermediate states obtained by individual rule applications are not
    counted. However, they can still be reported using the wrapper. As a
    result of counting only high level steps, this number indicates the amount
    of concurrency that takes place, since the difference in branching comes
    at synchronisation points only. 
  \item[Transitions.] The raw number of applications of individual rules.
    This includes rule applications that set the graph in a temporary state
    (i.e. a state that requires additional rule applications before it
    becomes a configuration as described above).
  \item[Initial Graph Size and Final Graph Size.] The sizes in terms of nodes and
    edges for each program. Since the translated programs do not depend on the
    \scoop semantics that is later applied, the initial graph sizes are the same
    across each semantics for a given program.
  \item[Time.] Wall clock time and standard deviation.
  \item[Memory.] Memory usage and standard deviation. Here, we report the
    peak amount of memory used by the Java VM executing the exploration
    process.
\end{description}

\tikzset{
    table/.style={
        matrix of nodes,
        row sep=-\pgflinewidth,
        column sep=-\pgflinewidth,
        nodes={
            rectangle,
            align=right
        },
        minimum height=1.5em,
        text depth=0.5ex,
        text height=2ex,
        nodes in empty cells,
        column 1/.style={
            nodes={text width=2em,text width=25ex,align=left}
        },
        column 2/.style={
            nodes={text width=2em,text width=8ex,align=center}
        },
        column 3/.style={
            nodes={text width=11ex}
        },
        column 4/.style={
            nodes={text width=10ex}
        },
        column 5/.style={
            nodes={text width=10ex}
        },
        column 6/.style={
            nodes={text width=10ex}
        },
        column 7/.style={
            nodes={text width=16ex}
        },
        column 8/.style={
            nodes={text width=12ex}
        },
        row 1/.style={
            nodes={
                fill=gray,
                text=white,
                font=\tiny\bfseries,
            }
        }
    }
}

\begin{table}
  \caption{Evaluation results (graph size and final graph size given as number of nodes/number of edges, time in seconds, memory in GB, and the latter two with standard deviation in seconds and GB respectively)}
  \label{tab:evaluation}
  \begin{adjustbox}{max width=\textwidth}
\begin{tikzpicture}

\matrix (first) [table]
{
Initial Graph & Runtime & Configurations & Transitions & Graph Size & Final Size & Time [std] & Memory [std]\\
                DP 2 eager (no commands) &      QoQ &   443    & 6135    & 254 / 395 & 300 / 473 &    5.380 [0.199]  &    0.582 [0.000]\\
                                         &       RQ &   442    & 6010    & 254 / 395 & 300 / 473 &    5.409 [0.082]  &    0.580 [0.000]\\
                                         &   DSCOOP &   1247   & 16313   & 254 / 395 & 304 / 477 &   12.968 [0.581]  &    0.684 [0.020]\\
                              DP 2 eager &      QoQ &   5863   & 75818   & 226 / 343 & 282 / 456 &   25.579 [0.862]  &    1.744 [0.019]\\
                                         &       RQ &   4219   & 54441   & 226 / 343 & 261 / 396 &   18.270 [0.657]  &    1.677 [0.092]\\
                                         &   DSCOOP &   13046  & 166399  & 226 / 343 & 265 / 400 &   52.979 [0.566]  &    2.647 [0.163]\\
                 DP 2 lazy (no commands) &      QoQ &   919    & 11935   & 250 / 387 & 296 / 465 &    9.664 [0.447]  &    0.644 [0.015]\\
                                         &       RQ &   868    & 11211   & 250 / 387 & 325 / 541 &    9.138 [0.624]  &    0.641 [0.012]\\
                                         &   DSCOOP &   2303   & 28676   & 250 / 387 & 331 / 560 &   21.411 [0.496]  &    1.020 [0.011]\\
                               DP 2 lazy &      QoQ &   9609   & 123583  & 221 / 334 & 256 / 387 &   40.891 [0.776]  &    2.447 [0.196]\\
                                         &       RQ &   5679   & 72692   & 221 / 334 & 288 / 470 &   23.548 [0.807]  &    1.971 [0.131]\\
                                         &   DSCOOP &   18874  & 237124  & 221 / 334 & 294 / 489 &   73.001 [0.890]  &    3.388 [0.214]\\
                DP 3 eager (no commands) &      QoQ &  3286   & 45152   & 254 / 395 & 316 / 499 &   35.986 [1.055]  &    1.529 [0.002]\\
                                         &       RQ &  3269   & 43967   & 254 / 395 & 316 / 499 &   35.124 [0.867]  &    1.728 [0.032]\\
                                         &   DSCOOP &  14867  & 192100  & 254 / 395 & 322 / 505 &  147.302 [6.960]  &    3.933 [0.202]\\
                              DP 3 eager &      QoQ &  227797 & 2924382 & 226 / 343 & 302 / 492 & 1480.638 [40.989] &   13.830 [0.241]\\
                                         &       RQ &  99198  & 1270216 & 226 / 343 & 277 / 422 &  436.354 [5.107]  &   11.491 [0.301]\\
                                         &   DSCOOP &  523513 & 6633232 & 226 / 343 & 283 / 428 & 2726.030 [40.534] &   13.785 [0.168]\\
                 DP 3 lazy (no commands) &      QoQ &  11774  & 151526  & 250 / 387 & 312 / 491 &  115.693 [3.137]  &    3.995 [0.032]\\
                                         &       RQ &  10877  & 139216  & 250 / 387 & 355 / 604 &  109.221 [2.352]  &    3.549 [0.088]\\
                                         &   DSCOOP &  47710  & 597564  & 250 / 387 & 364 / 632 &  474.863 [8.735]  &    7.896 [0.272]\\
                               DP 3 lazy &      QoQ &  444689 & 5684103 & 221 / 334 & 272 / 413 & 2424.935 [92.014] &   13.934 [0.067]\\
                                         &       RQ &  170249 & 2166740 & 221 / 334 & 319 / 536 & 1090.135 [29.512] &   13.887 [0.125]\\
                                         &   DSCOOP &  1288663 & 16176547 & 221 / 334 & 278 / 421 & 5999.547 [56.999] &   13.963 [0.188]\\
                              barbershop &      QoQ &  54325  & 702611  & 302 / 466 & 346 / 538 &  488.813 [2.994]  &    8.252 [0.142]\\
                                         &       RQ &  38509  & 494491  & 302 / 466 & 346 / 538 &  342.980 [3.825]  &    7.096 [0.244]\\
                                         &   DSCOOP &  179392 & 2270388 & 302 / 466 & 350 / 542 & 1954.988 [36.668] &   13.772 [0.071] \\
                                    PC 5 &      QoQ &  12366  & 156210  & 307 / 476 & 353 / 548 &  135.797 [4.408]  &    3.417 [0.110]\\
                                         &       RQ &  4085   & 51283   & 307 / 476 & 353 / 548 &   45.107 [2.377]  &    2.080 [0.137]\\
                                         &   DSCOOP &  23174  & 286641  & 307 / 476 & 356 / 551 &  246.470 [3.795]  &    5.201 [0.168]\\
                                   PC 20 &      QoQ &  50286  & 632820  & 307 / 476 & 398 / 593 &  575.061 [30.652] &    7.719 [0.353]\\
                                         &       RQ &  12890  & 159958  & 307 / 476 & 398 / 593 &  141.640 [3.734]  &    4.318 [0.098]\\
                                         &   DSCOOP &  90434  & 1113531 & 307 / 476 & 401 / 596 &  997.760 [27.277] &   10.961 [0.383]\\
                          dining savages &      QoQ &  79398  & 1008596 & 410 / 631 & 459 / 716 & 1240.665 [36.165] &   11.738 [0.397]\\
                                         &       RQ &  35361  & 448576  & 410 / 631 & 459 / 716 &  530.563 [24.885] &    7.120 [0.081]\\
                                         &   DSCOOP &  303678 & 3789448 & 410 / 631 & 473 / 751 & 5094.824 [35.232] &   13.925 [0.131]\\
};

\begin{pgfonlayer}{background}
  \foreach \x / \y in {5/7,11/13,17/19,23/25,29/31,35/37}
  \path[fill=gray!20] (first-\x-1.north west) rectangle (first-\y-8.south east);
\end{pgfonlayer}

\end{tikzpicture}
\end{adjustbox}
\end{table}

To obtain the results, we used the most recent version of our compiler and
wrapper~\cite{repo} at the time of writing. Furthermore, we used \groove
5.5.6, also the most recent version available. The time and memory values are the means of
five runs. All experiments were carried out on a notebook with an
Intel Core i7-4810MQ CPU and 16 GB of main memory. We used the OpenJDK 1.8
Java VM with the \texttt{-Xmx 14g} option.

\paragraph{Benchmark Results.}
The results of the evaluation are reported in Table~\ref{tab:evaluation}.
We performed full state-space exploration for all combinations of the programs and execution models.

Since initial graphs are completely independent of the chosen semantics, the initial graph sizes within each program are the same. The initial graph
sizes increase linearly with the size of the translated input program.
The final sizes of the graphs are, however, larger, since the graphs now contain
the dynamic part of the state, and its related components such as handlers, objects, and queues.
The final states of a given program also differ across the semantics due to the
different topologies and representations of queues, for example. In order to keep the graph
sizes down, we use ``garbage collection'' rules, which remove edges and nodes
that are no longer needed during execution (i.e.~the results of intermediate
computations).  However, note that we do not perform real garbage collection.
For example, unreachable objects are not removed, and the graph size increases
linearly with the number of created objects.

The number of configurations gives us an insight into how the different
semantics behave, since this column only counts proper steps in the
exploration. Differences between these numbers
arise from different branching at synchronisation points, thus the number of
configurations is an indicator as to how much concurrency the semantics allows.
However, it is important to note that it is not a simple matter of ``higher is
better''. When comparing \req and \qoq, we observe that \qoq produces more
configurations, agreeing with our intuition that \qoq allows more concurrency
(or, in the context of \scoopgts, more branching at synchronisation points).
However, we can also see that using \dscoop results in more configurations. In
this case, this is due to the fact that \dscoop is more complex due to the additional (pre)locking protocol on top of \qoq.

The time and memory columns show the raw power requirements of our toolchain. The number of configurations is, unsurprisingly,
particularly sensitive to programs with many handlers and only asynchronous
commands (\eg dining philosophers). Programs that also use synchronous queries
(\eg producer-consumer) scale much better, since queries force synchronisation once
they reach the front of the queue. We note again that our aim was to
facilitate automatic analyses of representative \scoop programs that covered
the different usages of the language mechanisms, rather than optimised
verification techniques for production-level software. The results suggest
that for this objective, on benchmarks of the size we considered, the toolchain scales well enough to be
practical.

\myparagraph{Error Rules / Discrepancies Detected.} In our evaluation of the
various dining philosophers implementations, we were able to detect that the
lazy implementation (Listing~\ref{listing:philosopher}) can result in deadlock
under the \req model, but not under \qoq or \dscoop.  This was achieved by
using error rules that match circular waiting dependencies (such as the one exemplified by Figure~\ref{fig:fac:deadlock_rq}). In case a
deadlock occurs that is not matched by these rules, we can still detect that
the execution is stuck and report a generic error, after which we manually
inspect the resulting configuration.  While such error rules are useful for
analysing \scoopgraphs in general, it is also useful to define rules that
match when certain program-specific properties hold. For example, if we take a
look at the eager implementation of the dining philosophers
(Listing~\ref{listing:philosopher}) and its executions under \req, \qoq, and
\dscoop, we find that the program cannot deadlock under any one of them. This does not
prove, however, that the implementation actually solves the dining
philosophers problem under all semantics. To check this, we defined an error
rule that matches if and only if two adjacent philosophers are in their
separate blocks at the same time, which is impossible if forks are treated as
locks (as they implicitly are under \req). This rule matches
only under the \qoq and \dscoop semantics, highlighting that under these
semantics, the program is no longer a solution to the dining philosophers
problem. (We remark that it can be ``ported'' to \qoq and \dscoop by replacing
the commands on forks with queries, which force the waiting.) 

To summarise, we can distinguish between two kinds of rules for detecting
errors and discrepancies: (i)~rules that match generic
criteria, but depend on the details of the execution model (e.g.~cyclic deadlock
conditions with locked handlers in the \req model); and (ii)~program-specific
rules that match conditions specific to the program that is simulated (e.g.~match when two adjacent philosophers are eating at the same time). By systematically defining combinations of these kinds of rules for our benchmark programs and execution models, our workbench can provide a richer comparison.  

\section{Related Work}
\label{sec:fac:related_work}

We briefly describe some related work closest to the overarching themes of our paper: frameworks for semantic analyses, \gts models for concurrent asynchronous programs, and verification techniques for \scoop.

\myparagraph{Frameworks for Semantic Analysis.}
The closest approach in spirit to ours is the work on the \KK framework~\cite{DBLP:conf/wrla/LucanuSR12,DBLP:journals/jlp/RosuS10}.
It consists of the \KK concurrent rewrite abstract machine and the \KK technique. One can think of \KK as domain specific language for implementing programming languages with a special focus on semantics. It was recently successfully applied to give comprehensive semantics to Java~\cite{KJavaPOPL2015} and JavaScript~\cite{park-stefanescu-rosu-2015-pldi}. Both \KK and our workbench have the same user group (programming language designers and researchers) and focus on formalising semantics and analysing programs based on this definition. We both have ``modularity'' as a principal goal, but in a contrasting sense: our modularity is in the form of a semantic plug-in mechanism for parameterising different model components (e.g.~storage, synchronisation, network topology), whereas \KK focuses on modularity with respect to language feature reuse. In contrast to our approach, \KK targets the whole language toolchain, including the possibility to define a language and automatically generate parsers and a runtime simulator for testing the formalisation. Based on the formal power of \maude's conditional rewriting logic, \KK also offers axiomatic models for formally reasoning about programs, and offers the possibility to define complex static semantic features, \eg advanced typing and meta-programming.

Despite having similar underlying theoretical power (\KK's rewriting is similar to ``jungle rewriting'' graph grammars~\cite{DBLP:conf/gg/SerbanutaR12}), \scoopgts models make the graph-like interdependencies between concurrently running handlers (or threads of execution) a first-class element of the model. This is an advantage for analyses of concurrent asynchronous programs, as many concurrency properties can straightforwardly be reduced to graph properties (\eg deadlocks as wait-cycles). Our explicit \gts model also allows us to compare program executions under different semantics, which is not a targeted feature of \KK. We also conjecture that our diagrammatic notations are easier for software engineers to grasp than purely algebraic and axiomatic formalisations.

\myparagraph{Semantic Analysis of Memory Models.}
Memory models are crucial for defining the correctness of concurrent shared memory platforms and programming languages. There is a large body of work targeting formalisations (\eg axiomatic models as in \cite{Mador-Haim:2012}, operational models as in \cite{NienhuisMS16}) and reasoning about these memory models' power (\eg \cite{Higham97definingand}). A recent axis of work, \eg in \cite{Wickerson-et_al17a,Mador-HaimAM10}, targets the generation of litmus tests that formalise the differences between memory models of the C language family (including GPU programming). In our formal setting of asynchronous distributed programs, \eg \scoop, which is guaranteed to be data race free, memory models do not play as prominent a role for program analysis. However, providing a hands-on notion of semantic differences via a set of example programs (\ie litmus tests) is close in spirit to our workbench's general goal of making semantics more accessible to the programmer.

\myparagraph{\gts Models for Concurrent Asynchronous Programs.}
Formalising and analysing concurrent object-oriented programs with \gts-based models is an emerging trend in software specification and analysis, especially for approaches rooted in practice. See \cite{Rensink10} for a good discussion---based on a lot of personal experience---on the general appropriateness of \gts for this task.

In recent decades, conditional rewriting logic has become a reference formalism for concurrency models; we refer to~\cite{DBLP:journals/tcs/Meseguer92} and its recent update~\cite{DBLP:journals/jlp/Meseguer12} for details. While having a comparable expressive power, our decision to use \gts and \groove as our state-space exploration tool led us to an easily accessible, generic, and parameterisable semantic model and toolchain that executes in acceptable time on our representative \scoop examples.

Closest to our \scoopgts model is the \textsc{Qdas} model presented in \cite{Geeraerts-HR13a}: an asynchronous, concurrent, waiting queue based \gts model with \emph{global memory}, for verifying programs written in Grand Central Dispatch \cite{GCD-Reference}. Despite the formal work, there is not yet a compiler for transforming Grand Central Dispatch programs into configurations for the \gts model. Furthermore, the model was not designed with modularity of semantic components in mind.

The Creol model of \cite{JohnsenOY06} focuses on asynchronous concurrent models but without more advanced remote calls via queues as needed for \scoop. Analysis of the model can be done via an implementation in \maude~\cite{JohnsenOA05}.

Several approaches exist for analysing programs based on the actor model~\cite{Agha86a}, \eg Erlang~\cite{Armstrong-VW96a} (see also the discussion in Section~\ref{sec:fac:genericmodel}). Most approaches rely on reasoning about a program's correctness on an abstract level, \eg as in \cite{Summers2016} or \cite{Desai14}, and do not focus on comparing executions under different semantics.

There are a number of \gts-based models for Java, but they only translate the code to a typed graph similar to the control-flow subgraph of \scoopgts~\cite{CorradiniDFR04,RensinkZ09}. A different approach is taken by~\cite{FerreiraFR07}, which abstracts a \gts-based model for concurrent object-oriented systems~\cite{FerreiraR05} to a finite state model that can be verified using the SPIN model checker.  However, despite the intention to build generic frameworks for the specification, analysis, and verification of object-oriented concurrent programs, \eg in \cite{DottiDFRRS05,ZambonR11}, there are no publicly available tools implementing this long-term goal that are powerful enough for \scoop.

\groove itself was already used for verifying concurrent distributed algorithms on an abstract \gts level~\cite{Ghamarian-MRZZ12a}, but not on an execution model level as in our approach. Similar in spirit are \gts based models for ad-hoc broadcasting networks, \eg in \cite{DelzannoSZ12}, which target complex dynamic topologies of the participating distributed processes but only provide high-level abstractions of the participating processes (\eg by state machines). However, our approach's generic topology abstraction could easily combine these sophisticated, dynamic communication networks with powerful low-level semantic models of the participating processes to gain a better understanding of distributed systems from the bottom-up.

\myparagraph{\scoop Analysis / Verification.}
Various analyses for \scoop programs have been proposed, including: using a \scoop virtual machine for checking temporal properties~\cite{Ostroff-THS09a}; checking Coffman's deadlock conditions using an abstract semantics~\cite{Caltais-Meyer16a};  and statically checking code annotated with locking orders for the absence of deadlock~\cite{West-NM10a}. In contrast to our work, these approaches are tied to particular (and now obsolete) execution models, and do not operate on (unannotated) source code.

The complexity of other semantic models of \scoop led to scalability issues when attempting to leverage existing analysis and verification tools. In~\cite{Brooke-PJ07a}, \scoop programs were hand-translated to models in the process algebra \textsc{Csp} to perform, for example, deadlock analysis; but the leading \textsc{Csp} tools at the time could not cope with these models and a new tool was purpose-built to analyse them (but is no longer available online). In a deadlock detection benchmark for the \maude formalisation of \scoop under \req~\cite{Morandi-SNM13a}, the tool was not able to give verification results in any reasonable time (\ie less than one day) even for simple programs like dining philosophers\footnote{From personal communication with the researchers behind this benchmark.}; our benchmarks compare quite favourably to this. Note that since our work focuses more on semantic modelling and comparisons than it does on the underlying model checking algorithms, we did not yet evaluate \groove's generic bounded model checking algorithms for temporal logic properties on our \scoopgts models.

\section{Conclusion}
\label{sec:fac:conclusion}

We proposed and constructed a semantics comparison workbench for \scoop, a concurrent, asynchronous, and distributed programming language based on message-passing, and used it to compare behavioural and safety properties of programs under different execution models. We constructed the workbench by applying the following general steps:
\begin{inparaenum}[(i)]
	\item derive a graph-based, compositional metamodel to which the family of execution models or semantics all conform;
	\item formalise the different semantics as \gts rules and control programs (strategies) in \groove, exploiting modularity and semantic parameterisation to obtain versatile and extensible models;
	\item test the formalisations by comparing simulations in \groove against the actual implementations;
	\item ensure soundness by evaluating the rules in expert interviews, and where possible, formally relating any existing semantics to the \gts rules and strategies;
	\item express generic safety properties (e.g.~``this will not deadlock'') and benchmark-specific properties (e.g.~``adjacent philosophers will not eat at the same time'') as special error rules, that match only when a state violates the property; 
	\item apply the \gts model checking engine of \groove to check whether error rules are applied consistently (or not) for a program under different semantics.
\end{inparaenum}

We presented a compositional semantics metamodel for \scoop, and used it to construct \scoopgts, a formalisation in \groove that covered the principal execution models of the language and a recent extension for distributed programming. We highlighted how common components could be used modularly across semantics, and how the components that differed (e.g.~request queues and synchronisation) could be ``plugged-in'' by exploiting the modelling power of \gts and control programs in \groove. We built a wrapper for \groove that automates the translation of \scoop source code into an initial configuration (i.e.~a graph), triggers its \gts state-space exploration algorithms, and reports to the user differences between the state-spaces under different semantics (e.g.~number of transitions, graph sizes) as well as any error rule applications detected. We applied the wrapper to a set of \scoop benchmark programs representing idiomatic usages of its abstractions, and detected a behavioural and deadlock-related discrepancy between the principal execution models, suggesting the usefulness of the workbench for comparing different semantics.

We are currently working on extending \scoopgts to cover some more advanced and esoteric features of \scoop and \dscoop (e.g.~exception handling~\cite{Morandi-Nanz-Meyer12a}, compensation~\cite{Schill-PM16a}, passive handlers~\cite{Morandi-NM14a}), and plan to extend the benchmark set to produce a comprehensive conformance test suite for the \scoop family of semantics. We are continuing to look for ways of refactoring \scoopgts to improve performance and broaden the class of programs it can handle practically, noting the impact that the shapes of rules and control programs can have on \groove's running time~\cite{Zambon-Rensink14a}. We also plan to explore the feasibility of applying formal \gts program logics (e.g.~\cite{Habel-Pennemann09a,Poskitt-Plump12a}) to our \groove models, in order to prove general properties of the execution models. Many properties of interest in \scoopgts involve arbitrarily long paths and cycles (e.g.~for defining general cyclic deadlocks), which require reasoning systems able to handle monadic second-order graph properties, e.g.~\cite{Poskitt-Plump14a}.

A more general line of future work would focus on the shape of \scoopgraphs in the state spaces generated by \scoopgts. Insights here would help us to devise better abstraction techniques (along the lines of~\cite{BackesR15}), which we could use to implement more efficient verification algorithms, and which could help us to visualise the influence of different semantic components on \scoopgraphs. Furthermore, we plan to build semantics comparison workbenches for other message-passing (or actor-like) concurrent and distributed programming languages in order to properly assess how effectively our approach generalises beyond \scoop. It would be particularly interesting if we could compare not only different execution models, but also different programming abstractions across multiple languages, all within one unified formalisation. \\


\begin{spacing}{0.5}
\myparagraph{Acknowledgements.} This work extends the research reported in our FASE 2016 paper~\cite{Corrodi-HP16a}, which was partially funded by ERC Grant CME \#291389.
\end{spacing}


\bibliographystyle{alpha}

\bibliography{references}

\end{document}